\documentclass[aps,reprint,prd]{revtex4-1}
\usepackage{graphicx} 
\usepackage{color} 
\usepackage{subfigure}
\usepackage{amsmath} 
\usepackage{amssymb}
\usepackage{bm}
\usepackage{exscale}
\usepackage[mathscr]{eucal}
\usepackage{upgreek}
\usepackage{color}
\usepackage{soul}

\begin{document}
\title{A Molecular Dynamics Study of The Translation and Rotation of Amphiphilic Janus Nanoparticles at a Vapor-Liquid Surface}

\author{Joel Koplik$^1$}
\email{jkoplik@ccny.cuny.edu}
\author{Charles Maldarelli$^2$}
\email{cmaldarelli@ccny.cuny.edu}
\affiliation{Benjamin Levich Institute and Departments of Chemical
Engineering$^2$ and Physics$^1$ \\
City College of the City University of New York, New York, NY 10031}

\date{\today}

\begin{abstract} 
We study the effects of heterogeneity on interfacial pinning and hydrodynamic 
drag using molecular dynamics (MD) simulations of Janus nanospheres at a
liquid/vapor interface.  We construct the free energy landscape for this
system, both in the continuum approximation using surfaces tensions and
the flat-interface approximation and atomistically using MD and thermodynamic 
integration. The results of the two methods differ in detail due to
interfacial distortion and finite width, as well as thermal fluctuations,
and only the MD landscape is consistent with simulations of a nanosphere 
approaching the interface from the liquid or vapor side.
When dragged along an interface, these Janus particles exhibit a
velocity-dependent tilt accompanied by a weak variation in drag force, but
never an enhancement of the drag force beyond the value when fully immersed.
This velocity dependence arises when the interface is pinned at
heterogeneities and prevents the particle from rotating, and 
similar behavior is observed for homogeneous but non-spherical particles.
The occurrence of different particle orientations having different drag 
coefficients may lead to an apparent violation of the Stokes-Einstein relation.
\end{abstract} 

\maketitle 

\section{Introduction}
\label{intro}
Dating back to the century-old publications of Ramdsen\cite{Ramsden03} and Pickering\cite{pickering06}, it has long been recognized that colloids (particles order  nanometers to microns in characteristic size)  can adsorb strongly and irreversibly to an apolar/polar fluid interface (e.g. an air/water or oil/water interfaces) to form monolayers. Over the past century, sustained interest in these monolayers derives from their ability, when adsorbed at the bubble interfaces of foams or the droplet interfaces of emulsions, to stabilize these dispersions from  coalescence through steric repulsion\cite{Binks2002, Binks06}. The driving force for the adsorption and entrapment of the colloid at the interface is a reduction in interfacial free energy;  upon breaching a fluid interface the colloid removes fluid interfacial area and exchanges the fluid/solid particle surface energy of one of the phases bounding the interface with the fluid/solid energy of the opposite phase. While colloidal monolayers provide an interfacial armor which stabilizes foams and emulsions, these  dispersions are not thermodynamically stable due to the fraction of the fluid interface that is not covered by particles\cite{aveyard2003emulsions, aveyard2012can}.

To reduce the interfacial energy of a surface particle further, attention over the past few  decades has focused on  amphiphilic spherical Janus colloids\cite{casagrande1989janus, ondarccuhu1990specific, binks2001particles, jiang2007janus,jiang2008controlling}. These are heterogeneous particles with two faces that have different wettabilities (contact angles) to the bounding phases, i.e one apolar surface which preferentially wets the apolar phase and the other  which preferentially wets the polar phase. We focus here  on the class in which the apolar and polar faces are hemispheres. 
Calculations done in the absence of  interface deformation detail an energy landscape  with a deep minimum when the contact line is pinned at the Janus boundary ( the preferred ``Janus configuration'',
and no equilibrium angle conditions are satisfied\cite{rezvantalab2013capillary, xie2015tunable, wang2016wetting}. Due to the Janus amphiphilicity, this minimum -  which can exceed the  minimum for a homogeneous particle with the same radius and contact angle by a factor of three -  demonstrates the potential of Janus colloids to form thermodynamically stable emulsions\cite{aveyard2012can} and continues to drive  research in the use of Janus particles to stabilize foams and emulsions .

The elaboration of techniques to functionalize the polar and apolar surfaces of Janus particles with functionalized chemistries has driven research into applications of these particles at fluid interfaces beyond dispersion stabilization, for reviews\cite{kumar2013amphiphilic,fernandez2016surface,  bradley2017janus}.  As  amphiphilic Janus particles are trapped at the interface, and their energy minimum sets a preferred configuration in which the  apolar and polar sides of the Janus particle face the apolar and polar phases,  monolayer assemblies of these particles - particularly closed packed arrangements -  can be used as the starting points for the fabrication of surface coatings or materials with controlled anisotropy (for example, surfaces with optofluidic mirrors, \cite{bucaro2008tunable}).  Advances in functionalization have also made possible the fabrication of active Janus colloids, in which one face is functionalized with a catalyst which mediates the reaction (typically in an aqueous phase)  of  a solute to a product, and the asymmetric distribution of reactant and product across the colloid creates a force imbalance on the particle which propels the particle \cite{sanchez2015, Dey2016, ebbens2016, bechinger2016,zottl2016, sen2017, Posner-rev}. If the opposite side of the Janus motor is less wettable than the reactive side to water, the locomotor can become trapped at an air/water or oil/water interface in the preferred Janus configuration, with the reactive side still in contact with the fuel and able to  drive the particle along the surface\cite{BTPSDS2012, wang2015enhanced, wang2017janus, isa2017}. The confinement at the interface provides a prescribed avenue for the locomotor to navigate in applications such as shuttling cargo or creating advection, and the orientation provides the accessibility of the active area to the fuel to sustain the motion.

These more recent applications of Janus particles at fluid interfaces center around the motion of the colloid along the surface, either as they self-propel or self-assemble, due to overlap of menisci around the particles created when the interface attaches to the Janus boundary, as for example when the Janus boundary causes an undulating contact lines \cite{Kralchevsky:2000mb, Stamou2000, brugarolas2011generation,park2011janus}, or  magnetized Janus particles are torqued by an external magnetic field to lift the Janus boundary\cite{xie2015tunable,xie2016controlled,xie2017direct}.   Whatever the origin, the movement of a Janus particle along a fluid interface with velocity $U$ generates hydrodynamic shear stresses on the particle which can be asymmetric relative to the interface\cite{dorr2015, cappelli2015dynamic}. If, for example, the viscosity of the apolar phase is smaller than the polar phase, the translation generates a greater shear stress in the polar phase creating a net  torque which acts to rotate the colloid counter-clockwise.
Rotational motion forces the polar phase at the contact line boundary to recede  over one side of the particle surface and advance over the surface on the opposite side. This contact line motion has been extensively studied for liquid droplets moving over planar substrates\cite{Leger:1992fr,Quere:2008dk}, and it is well recognized that topological and chemical heterogeneities on the substrate surface  can pin the contact line until hydrodynamic stresses are large enough  to cause the contact line to advance or recede. The surfaces of  non-Janus colloidal particles are typically heterogeneous. Measurements of the movement of these colloids toward a fluid interface\cite{kaz2012physical} show an extended time for the colloid to settle to its equilibrium configuration, and the measurements of the contact angles of liquids on colloidal particle surfaces are not very reproducible\cite{Rubio2014}. Both phenomena indicate that the contact line tends to become pinned on the particle surface and only advances slowly to an equilibrium position. Janus particles will exhibit two levels of surface heterogeneity; one, the heterogeneity of the individual phases and second, the potentially significant heterogeneity at the Janus boundary which is chemical and can be topological depending on how the particle functionalization is undertaken (e.g. if the Janus face is fabricated by sputter coating metallic layers). When a Janus particle begins to translate along the surface in its preferred configuration the hydrodynamic torque due to the translation can be resisted by the contact line becoming pinned at the Janus boundary, if the particle velocity and associated hydrodynamic torque is not large enough to allow the contact line to move. 

The issue of whether the Janus particle rotates as it moves, or remains in a fixed orientation, is central to many emerging applications of Janus particles at fluid interfaces, as the particles are required to remain in a fixed orientation (e.g. colloidal coatings and surface locomotors). The aim of this study is to examine whether the Janus boundary can pin the contact line of a Janus particle moving along the surface to prevent its rotation, and establish, if pinned, a relationship between the angle of arrest as a function of the translational velocity, and the resultant drag coefficient. To undertake this study, we use molecular dynamics (MD) simulations that can straightforwardly describe meniscus motion over topologically rough and chemically heterogeneous surfaces of nano-sized colloids. An additional advantage of using MD is that the interfacial free energy landscape of the Janus particles can be calculated through thermodynamic integration of forces at fixed Janus orientations through the interface, without the usual assumption that the interface is flat and does not equilibrate with the particle surface at the contact line. 

Using Lennard-Jones (LJ) interaction potentials, several MD studies\cite{Cheung2010,cheung2011, grest2012, rezvantalab2015molecular,dai2009a}, including our prior study\cite{koplik2017} (referred to below as KM), examined the surface diffusion of smooth and rough non-Janus) spherical colloids trapped and diffusing along  a liquid/liquid and a vapor/liquid interface. These studies demonstrated that the finite thickness of the interfaces in the MD simulations increased the diffusivity by decreasing the drag and that, for a vapor/liquid interface, the greater the immersion into the liquid phase, the smaller is the diffusion coefficient. The latter result was attributed to an increase in the viscous drag on the colloid in accordance with the Stokes-Einstein relationship and in agreement with continuum results \cite{fischer2006viscous, pozrikidis2007particle, danov2000viscous, dani2015hydrodynamics,dorr2016drag}. The effect of topological heterogeneities was examined in our prior study (KM) and Shojaei-Zadeh et al.\cite{rezvantalab2015molecular} by constructing atomistic colloids from a lattice assembles of nanoparticles, whereby the colloid is comprised of all particles in the lattice within a particular radius which then has a natural level of surface roughness. These rough colloids were found to rotate as they diffused with a stick-slip motion along the surface, indicating that the roughness was not sufficient to pin the interface when the particles moved by surface diffusion. KM also studied the steady translation of a colloid with a simulated rough surface along a vapor/liquid under a constant velocity. Under this circumstance, the colloids rotated and the computed drag coefficients were slightly less than continuum calculations of the drag \cite{fischer2006viscous, pozrikidis2007particle, danov2000viscous, dani2015hydrodynamics, dorr2016drag} due to the finite thickness of the interfacial layer. 

Very few MD simulations of the motion of Janus particles along fluid interfaces have been undertaken. Using LJ potentials, the diffusion of a Janus colloid has been simulated at a liquid-liquid interface for atomistic rough colloids \cite{gao2014orientation, rezvantalab2015molecular}. The Janus faces are constructed by assigning different interaction potentials for the particles comprising either half of the Janus particle. Both studies indicated that the colloids diffused along the surface in the Janus preferred configuration, providing the first indication that the Janus boundary pins the interface for diffusional motion (see also the Monte-Carlo simulations\cite{cheung2009stability} which come to the same conclusion). Simulations of the translation of a Janus colloid along the surface due to an applied force have not been undertaken. However, in a related study Shojaei-Zadeh et al\cite{rezvantalab2016shear,rezvantalab2016tilting} demonstrated that if the interface is sheared perpendicular to its normal, the resultant torque due to the shear rotated a Janus colloid from its preferred configuration with the contact line pinned at the Janus boundary. The rotation continued until the capillary torque derived from the pinning arrested the rotation.
\begin{figure}
\centering
\includegraphics[width=0.4\textwidth]{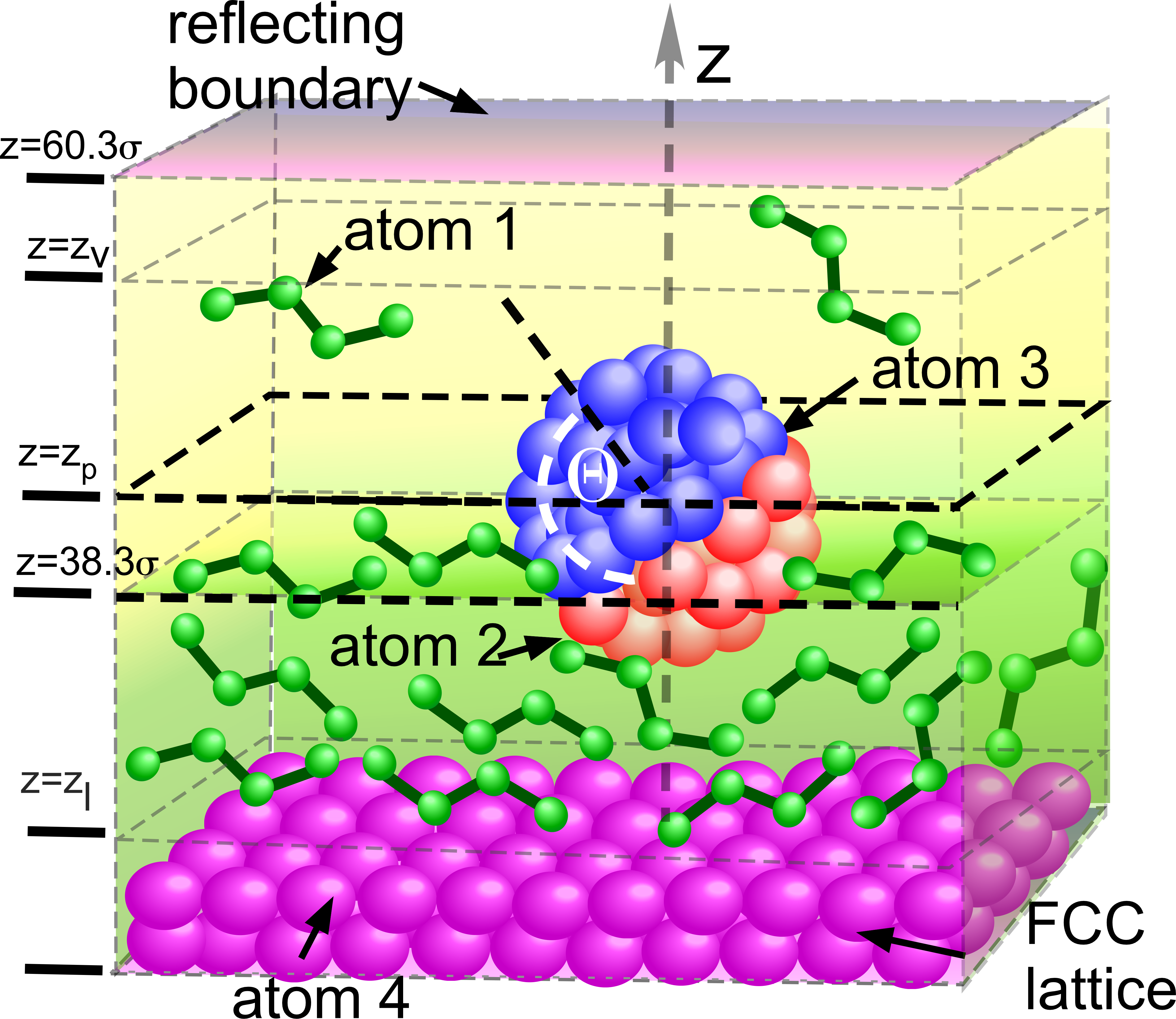}
\caption{Computational cell showing the tetrameric liquid (``1''), the FCC lattice base (``4'') and the Janus nanoparticle consisting of ``2'' and ``3'' at the fluid interface.}
\label{setupa}
\end{figure}

We begin by describing the formulation of our MD simulations, which employ LJ potentials. We study a Janus nanoparticle with a topologically rough surface attached to a vapor/liquid interface, with one hemisphere preferentially wetting the liquid phase relative to the opposite phase. We first compute the interfacial energy landscape accounting for the formation of a meniscus around the particle, and we use these calculations to understand how Janus colloids orient as they adsorb from either the vapor or liquid phase. We next examine their translation and rotation along the interface, where we focus on whether the interface can become pinned at the Janus boundary to prevent rotation and the effect of the suppression of the rotation on the hydrodynamic drag, and follow with conclusions.

\section{Formulation}
\label{Formulation}
The computational domain for the MD simulations is shown in Fig.~\ref{setupa}, and follows KM\cite{koplik2017}.  Tetramers of Lennard-Jones (LJ) atoms (species 1) tethered by springs comprise a molecular fluid which fills  the bottom of the simulation box, with a solid bounding surface at the bottom (made of LJ atoms (species 4) tethered in two layers to a fcc lattice configuration), and a (very low density) vapor fills the remainder. A reflecting wall is at the top of the box and  periodic  boundary conditions are imposed in both horizontal directions.   All atoms in the system interact via an adjustable Lennard-Jones potential $\displaystyle{ V_{\rm LJ}(r) = 4\,\epsilon\, \left[ (r/\sigma)^{-12} - c_{ij}\, (r/\sigma)^{-6}\ \right]}$, involving an atomic diameter $\sigma$ and an energy scale $\epsilon$, and the parameter $c_{ij}$ controls the strength of the interaction between atoms of species $i$ and $j$. All atoms have a common mass $m$, and we define a time scale $\tau=\sigma(m/\epsilon)^{1/2}$. The tetramer molecular chain is tied together by adding a FENE potential $V_{\rm FENE}=-{\frac{1}{2}}k_{F}r_0^2\log{(1-{\bf r}^2/r_0^2)}$ between adjacent atoms along the chain, with parameters ($k_{F}=30 \epsilon/\sigma^{2}, r_{o}=1.5 \sigma$) taken from \cite{grest1986}.  This molecular liquid is preferable to a monatomic one in this situation  because it has a sharper interface.  The simulation box is a cube of side 60.3$\sigma$, and the temperature is fixed at $0.8\epsilon/k_{B}$ using a local Nos\'e-Hoover thermostat. The MD simulations follow standard methods\cite{frenkelschmidt,Allantildesley}. The interactions between two liquid atoms and between liquid and the solid base and the solid base atoms have the standard value $c_{11}=c_{14}=c_{44}=1$. Under these conditions the liquid density is 0.857$\sigma^{-3}$ and separate simulations of Couette flow of the tetramer demonstrates Newtonian behavior with a viscosity  equal to $\eta=5.18m/(\sigma \tau)$. The surface tension $\gamma^{f}$, obtained as the difference in the normal and transverse stress across the interface is 0.668$\epsilon/\sigma^{2}$. The thickness of the interfacial zone is approximately 3$\sigma$ and in the absence of the particle is located at  $38.3 \sigma$ as measured from the bottom of the FCC lattice

\begin{figure*} 
\centering
\subfigure[]{\includegraphics[width=0.35\textwidth]{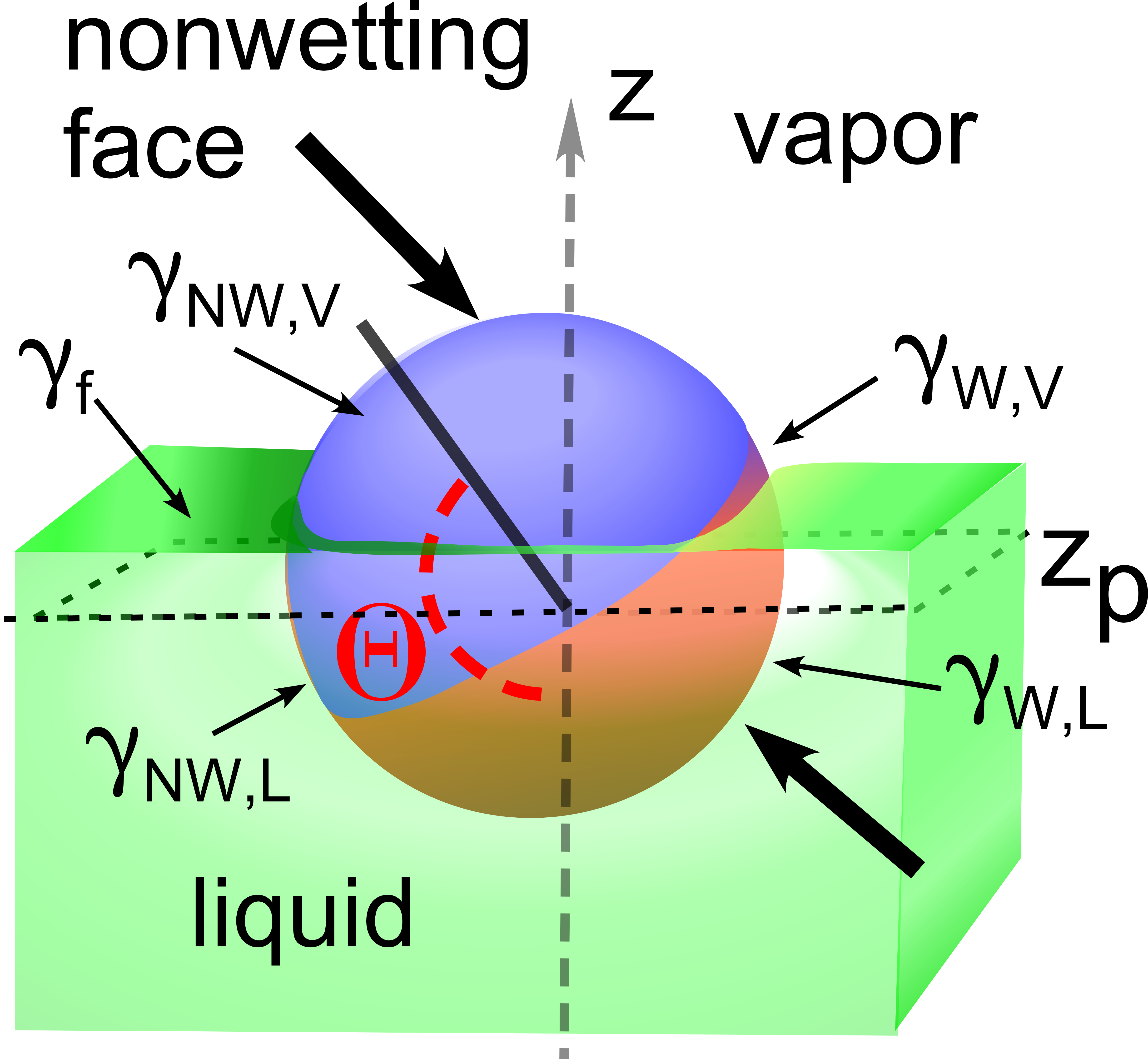}}\hspace{0.25in}
\subfigure[]{\includegraphics[width=0.5\textwidth]{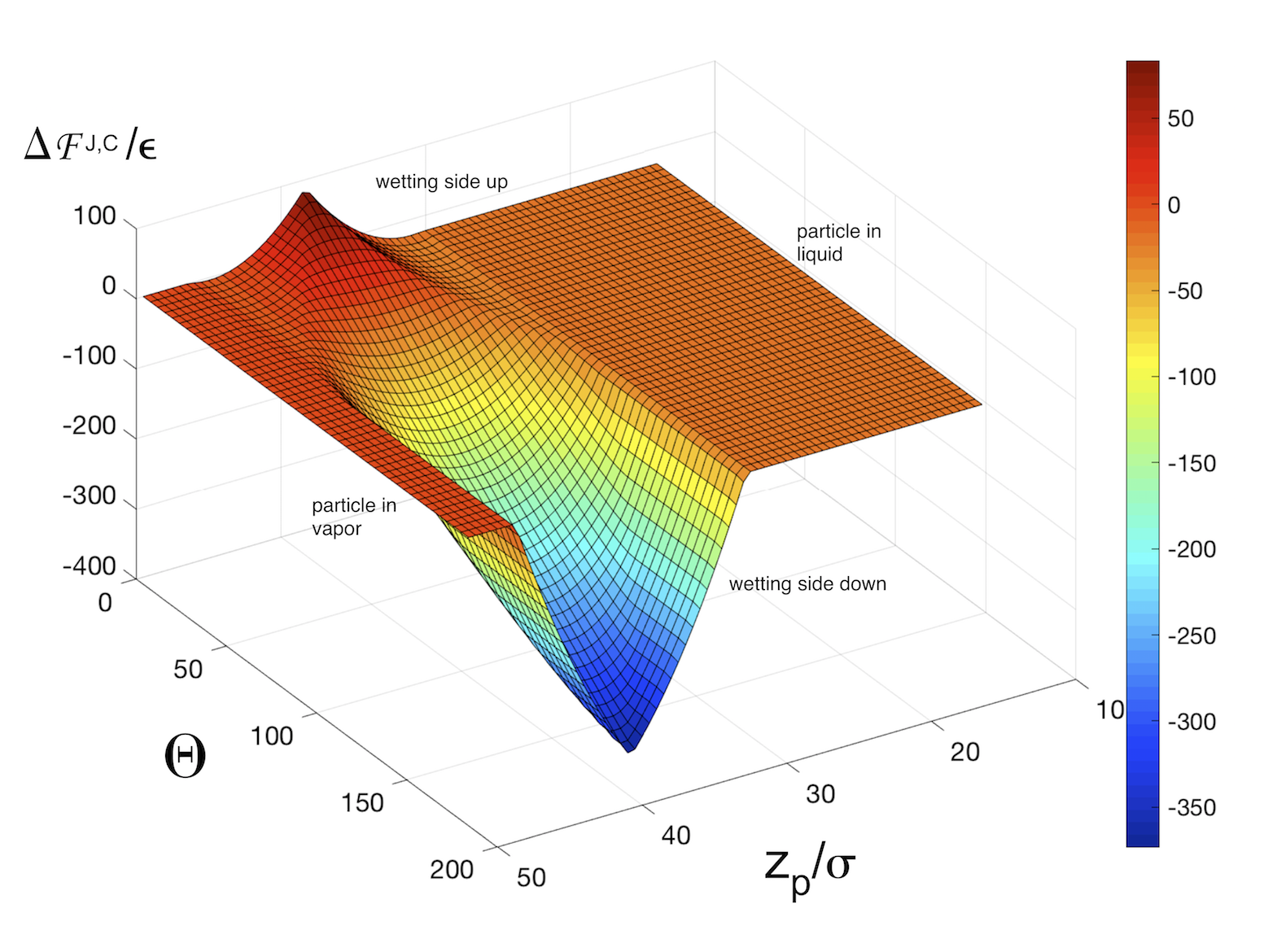}}
\caption{Continuum calculation of the free energy landscape  of a Janus particle as a function of orientation $\theta$ and position $z_p$ from the solid free energies of the Janus faces, assuming a flat fluid interface. (a) geometry of the calculation, (b) free energy surface for a 1.2/0.7 Janus particles.}
\label{continuumenergy}
\end{figure*}

As described above and following our prior study, the particle is constructed by cutting out a spherical section (radius $8\sigma$) of a cubic lattice of LJ atoms with the same density as the liquid (0.857 $\sigma^{-3}$) to create a composite with a resulting surface roughness, and we assign different interaction strengths of the atoms in the two halves of the particle with the liquid tetramer atoms  to construct a Janus particle. The atoms in the two hemispheres are denoted as species $2$ and $3$ (Fig.~\ref{setupa}a) and the interactions $c_{12}$ and $c_{13}$ are varied to control the wettability of the faces of the Janus particle and hence its orientation and immersion depth into the liquid. For the strongly-wetting hemisphere we fix $c_{12}=1.2$ and for the other hemisphere either $c_{13}$=0.7 or 0.8.  As we demonstrated in KM, a uniform sphere with $c_{13}=1.2$ would be completely immersed in the liquid in equilibrium. A uniform particle with $c_{13}$=0.8 and 0.7 straddle the interface with apparent contact angles equal to 111$^{\circ}$ and 132$^{\circ}$, respectively\cite{koplik2017}, indicating that the particle situates with its center above the free interface. The $c_{13}$ particle situates at a higher position due to its smaller interaction with the liquid. If the same interactions were used for a spreading drop on an planar atomic surface, a $c_{13}=1.2$ liquid would spread completely while a $c_{13}=0.8$ liquid would form a partially wetting drop. For convenience, we will denote the Janus particles by their interactions, e.g. $c_{12}/~c_{13}$.

\section{The Interfacial Energy Landscape of a Janus Particle}
\label{EnergyLandscape}
\subsection{Surface tensions and thermodynamic integration}

The \textit{continuum} value of the free energy of a particle is basically the integral over the surface area of the appropriate local free energy per area -- the surface tension. 
As noted earlier, the common simplifying approximation is to assume that the 
interface is flat up to the particle contact-line\cite{rezvantalab2013capillary, xie2015tunable, wang2016wetting}, thereby avoiding the difficult calculation of the area of a distorted interface. We consider Janus particle at an interface separating an apolar/polar liquid phase, or equivalently here a vapor/liquid interface, as shown Fig. \ref{continuumenergy}a. For each orientation angle $\theta$ and immersion depth $z_{p}$, the particle interfacial free energy is computed by first integrating over each point $\bm{r}$ on the particle surface a differential surface area element $dS(\bm{r})$ times one of the Janus particle's solid energies per unit area, $\gamma_{i,j}$ where $i$ indicates the face ($i$=NW or W for nonwetting or wetting) and $j$  denotes the phase attached to the interface ($j$=V or L for vapor or liquid). The energy used at $\bm{r}$ depends on whether the surface point is within the polar or apolar section, and whether it is in contact with the apolar or polar phase. We then subtract from this integral the free energy of the circular region of the liquid/vapor interface (of radius $r_{S}$) removed by the presence of the particle: its area times the liquid/vapor surface tension.  Explicitly, 
the continuum interfacial energy at a position $z_p$ and angle $\theta$ relative to the energy when the Janus colloid is all in the vapor (apolar) phase is:
\begin{equation}
\Delta{\cal{F}}^{J,C}(z_p,\theta)= \int_{IS(r_p)}dS({\bm r}) [\gamma_{SL}({\bm r})-\gamma_{SV}({\bm r})] -\pi\,r_S^2\gamma^{f}
\label{fe1}
\end{equation}
where the superscript indicates the continuum energy calculation (C) of a Janus particle (J) with an assumed  flat interface,  and the integral runs over the (smooth) spherical immersed surface (IS) of the particle, which depends on $r_p$. A key point here is 
that while the absolute free energies of solid interfaces are difficult to calculate\cite{frenkelschmidt}, the free energy change due to replacing a surrounding vapor by liquid is more accessible. In the integral,  
$\gamma_{SL}(\bm{r})-\gamma_{SV}(\bm{r})$ represents either the difference
$\gamma_{NW,L}-\gamma_{NW,V}$ if at the position $\bf{r}$ the nonwetting
face is in contact with  the liquid  or $\gamma_{W,L}-\gamma_{W,V}$ if the
wetting  face is touching the liquid (polar phase). These solid surface
energy differences can be calculated by thermodynamic integration, which we
also use below for computing the molecular dynamics Janus particle energy
landscape for any orientation and immersion depth, and for an interface that may not be flat.

In the thermodynamic integration method, a path is selected for the configuration of the particle, the position of the center and the particle orientation in general, and at each configuration the particle is fixed and the system is first allowed to equilibrate and then the force on the particle is measured as a time average.  If the temperature is constant along the path ($\xi$), the change in free energy between any two configurations is the work done to move the particle reversibly between them. 
\begin{equation}
\Delta{{\cal{F}}^{J,MD}}|_{1-2} = -\int_1^2 d\xi\, F^{J}(\xi)
\label{fe2}
\end{equation}
At the moment we require only the local differences in surface tension, $\gamma_{NW,L}-\gamma_{NW,V}$ and so on, which is the difference between the free energies of homogeneous particles of that wettability exposed to
liquid or vapor, divided by the particle's surface area. 
The orientation is not relevant here, and we choose vertical paths normal to the bottom of the rectangular computational cell where the particle begins far above the interface at $z_p=z_V$ and ends far below it at $z_p=z_L$.
In the present system, the computational cell has $0\le z\le 60.3 \sigma$,
the average plane of the interface is located at $z$=38.3$\sigma$ (with a
thickness of approximately 3$\sigma$), and the particle has radius
8$\sigma$, so we choose $z_V$=$50\sigma$ and $z_L$=$15\sigma$ (if $z_L$ is too small the particle interacts with the bottom wall).  Then
\begin{equation}
\gamma_{i,L}-\gamma_{i,V}=-\frac{1}{4\pi R^2}\int_{z_V}^{z_L}dz\, F_i(z)
\end{equation}
where $F_i(z)$ is the force on a homogeneous particle with the appropriate solid liquid interaction interaction strength $c_{1i}$ when its center is at $z$.
Note that in this quasi-static integration, all interfacial energy effects are in principal included and, importantly, the interface shape is not fixed to a planar configuration but allowed to come to an equilibrium dictated by the orientation and position of the Janus particle. We note that along the surface but away from the particle the interface does become planar but for positions in which the particle breaches the interface (as we discuss in detail below) the interface is deformed in the immediate vicinity of the contact line as the fluid equilibrates with the solid energies.  The results are 
$\gamma_{W,L}-\gamma_{W,V} =-0.609\epsilon/\sigma^{2}$ for $c_{12}$=1.2, 
and $\gamma_{NW,L}-\gamma_{NW,V}$ equal to 0.313$\epsilon/\sigma^{2}$ for $c_{13}=0.8$ and 0.554$\epsilon/\sigma^{2}$ for $c_{13}=0.7$. Unsurprisingly, the stronger the interaction of the atoms of the colloid with the tetramer the more reduced the colloid energy relative to the gas phase.

\begin{figure}
\centering
\includegraphics[width=0.5\textwidth]{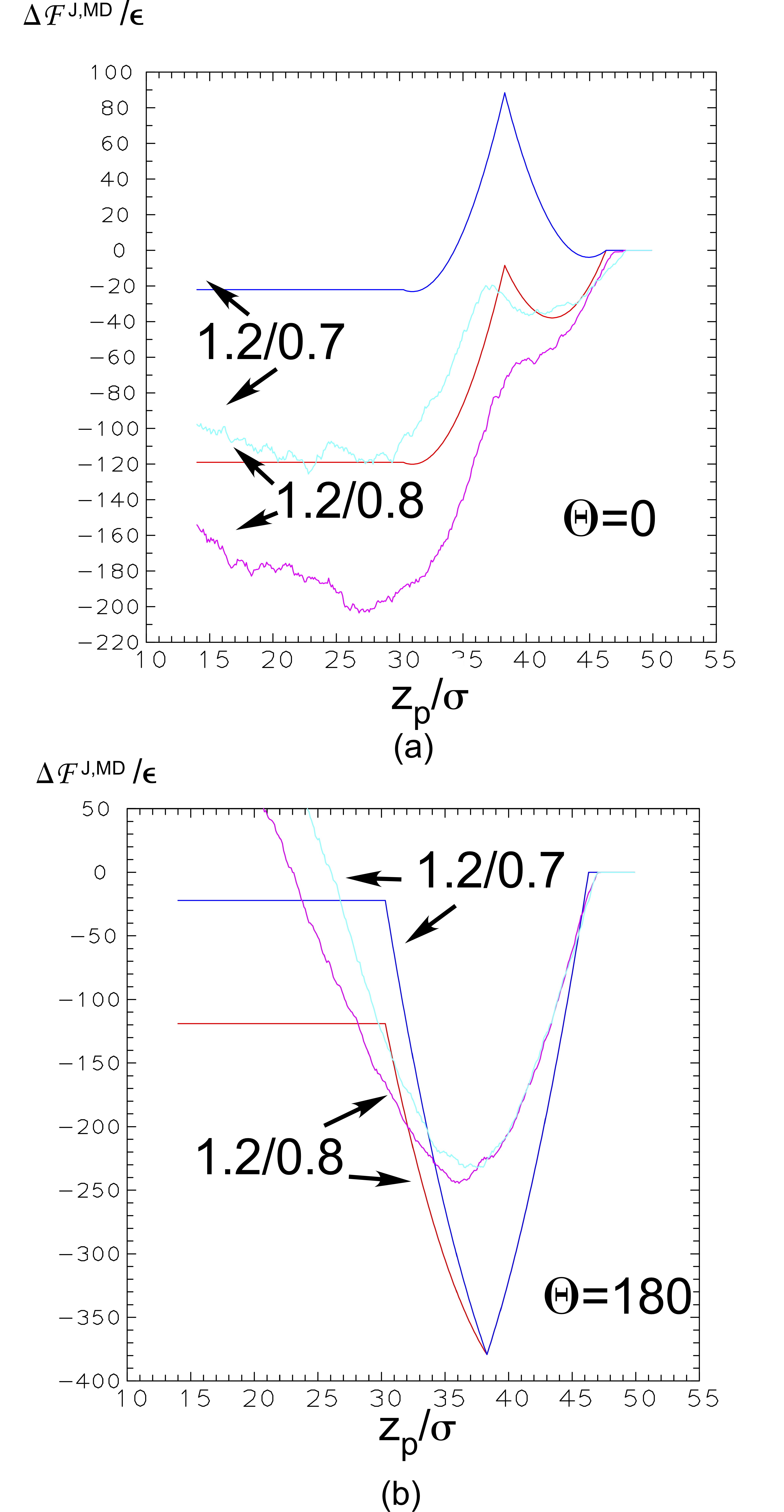}
\caption{Continuum and MD landscape at the fixed orientations along $\theta=0$ and $\theta=180^{\circ}$ for the 1.2/0.7 and 1.2/0.8 Janus particles.}
\label{onedlandscape}
\end{figure}

\subsection{Continuum landscape}

With the solid energy differences known, we can obtain $\Delta{\cal{F}}^{J,C}(z_p,\theta)$, the continuum landscape of a Janus particle (radius $R=8\sigma$) with an assumed flat surface (located at $z/\sigma = 38.3$), as a function of immersion depth ($z_{p}$) and orientation ($\theta$), by evaluating Eq.~\ref{fe1} numerically  for each value of $z_{p}$ and $\theta$. The results for the 1.2/0.7 particles is shown in Fig.~\ref{continuumenergy}b; the 1.2/0.8 landscape is very similar. There is sharp, global minimum at $\theta =180^{\circ}$ and $z_{p}/\sigma = 38.3$, which corresponds to the preferred Janus configuration with the wetting side down and fully immersed in the liquid phase. The value of this minimum is equal to $2 \pi R^{2} (\gamma_{W,L}-\gamma_{W,V})-\pi R^{2} \gamma_{f}=-379.2 \epsilon$ since the wetting hemispherical face is now completely immersed in the liquid.  More insight into this minimum and the nature of the landscape in general can be gained by examining the behavior of the interfacial free energy as a function of $z_p$ for particular orientations $\theta=180^{\circ}$ and  $\theta=0^{\circ}$ (Fig. \ref{onedlandscape}). At $\theta=180^{\circ}$, $\Delta{\cal{F}}^{J,C}$ has
a single minimum at $z_p/\sigma = 38.3$ which is easily rationalized: as the colloid, breaches the interface from the vapor with its wetting side down and becomes immersed in the liquid, the favorable (negative free energy relative to the wetting of this face with vapor) interaction of the wetting face with the liquid together with the decrease in the area of the fluid interface results in the monotonic reduction in the free energy to the minimum at the preferred Janus configuration when $z_p/\sigma=38.3$. As the particle immerses further into the liquid, the nonwetting face comes in contact with the liquid, and the unfavorable (positive free energy) interaction and the increase in the fluid area results in the monotonic increase in the free energy until the asymptotic value $\Delta{\cal{F}}^{J,C} = 180^\circ = 2 \pi R^{2} \left[(\gamma_{W,L}-\gamma_{W,V})+(\gamma_{NW,L}-\gamma_{NW,V}) \right]$ = $-22.1 \epsilon$ is achieved when the colloid is fully immersed in the liquid phase at $z_p/\sigma$ = $30.3$. 

For immersion of the colloid into the liquid from the vapor side with the wetting side up ($\theta=0^{\circ}$), two minima are observed due to a competition between the changes in the nonwetting and wetting solid surface energies and the fluid interface energy. As the colloid first breaches the interface from the vapor nonwetting side down, the reduction in the liquid interfacial area decreases $\Delta{\cal{F}}^{J,C}$, while the fact that the nonwetting face is coming in contact with the liquid increases the free energy.  This competition gives rise to the first minimum, and is the same competition which occurs when homogeneous nonwetting colloids (contact angles greater than 90$^{\circ}$) breach the interface of the liquid from the vapor side.  In this case the interfacial energy is given by 
\begin{align}
\begin{split}
 & \Delta {\cal{F}}^{J,C}(d+38.3\sigma,0^\circ )   =  \\
& \pi {R^2}\left\{ {2\left[ {{\gamma _{NW,L}} - {\gamma _{NW,V}}} \right]\left[ {1 - \frac{d}{R}} \right] + {\gamma _f}\left[ {\frac{{{d^2}}}{{{R^2}}} - 1} \right]} \right\}
\end{split}
\label{fe4}
\end{align}
where $d$ is the distance from the particle center in the vapor to the interface. The minimum is at $d_{min}/R = (\gamma_{NW,L} - \gamma_{NW,V})/\gamma_f$ 
or $z_p=38.3\sigma+d_{min}=44.9\sigma$ using the parameters for 1.2/0.7, which is the observed location of the first minimum. Thus the location of the first minimum is the equilibrium position that would be achieved by a homogeneous particle with an interaction coefficient with the solvent tetramer equal to $0.7$.  With continued immersion the free energy achieves a maximum when the particle center reaches the surface ($z_{p}/\sigma = 38.3$) with energy $2 \pi R^{2} (\gamma_{NW,L}-\gamma_{NW,V})-\pi R^{2} \gamma_{f}=88.5 \epsilon$.  This positive value reflects the large energy cost is immersing the nonwetting side completely into the liquid phase and represents globally the largest value for the free energy as the nonwetting face is completely wet by the liquid. With further immersion into the liquid, the wetting face comes into contact with the liquid, and this favorable (negative) interaction energy dominates the loss in the area of the vapor/liquid interface and accounts for a sharp decrease in $\Delta{\cal{F}}^{J,C}$ for $z/\sigma <  38.3$. Finally, just before the colloid becomes completely immersed, a second minimum arises which represents the minimum in the free energy of a homogeneous colloid with an interaction coefficient of 1.2 with the tetramer solvent which occurs at a position $z_p=38.3\sigma + R(\gamma_{W,L} - \gamma_{W,V})/\gamma_f = 31.0\sigma$  as observed in Fig.~\ref{continuumenergy}. With further immersion, the free energy increases and becomes equal to the completely immersed value  $2 \pi R^{2} \left [(\gamma_{W,L}-\gamma_{W,V})+(\gamma_{NW,L}-\gamma_{NW,V}) \right ] = -22.1 \epsilon$. 

\begin{figure*}
\centering
\subfigure[]{\includegraphics[width=0.6\textwidth]{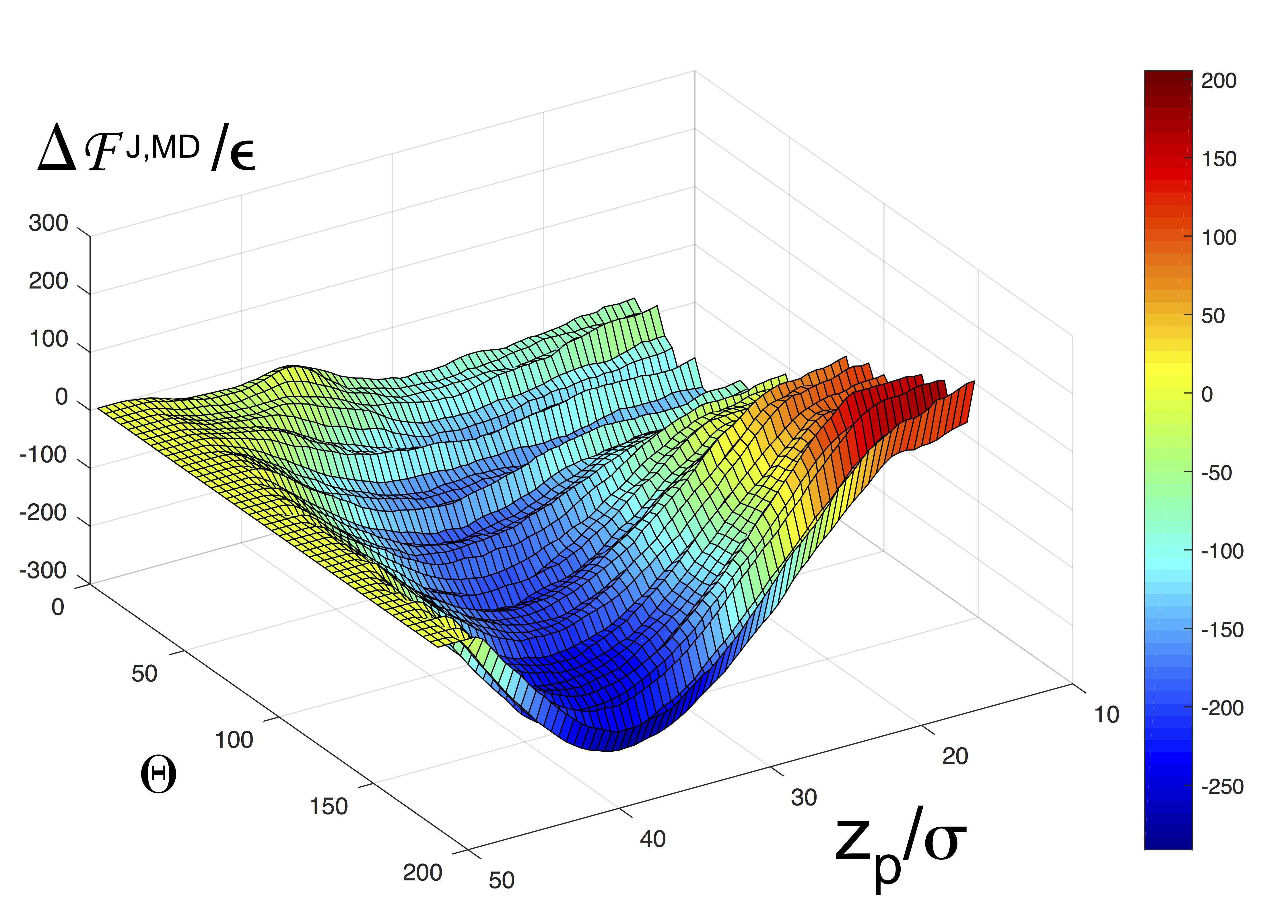}}
\subfigure[]{\includegraphics[width=.7\textwidth]{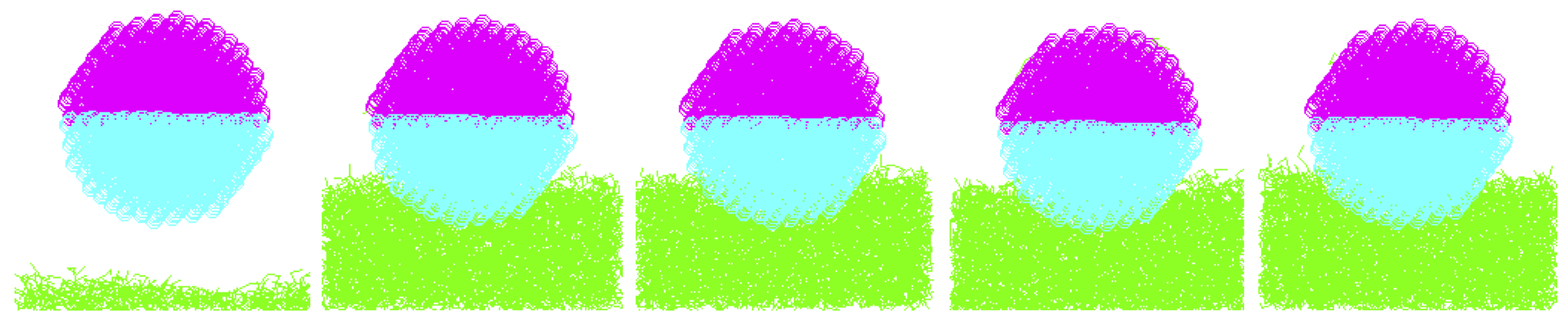}}
\subfigure[]{\includegraphics[width=.7\textwidth]{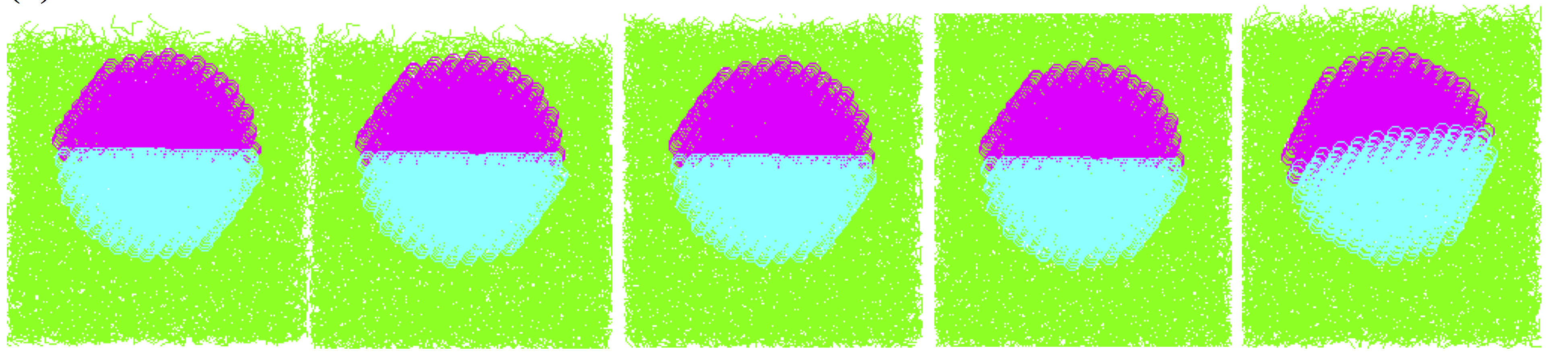}}
\subfigure[]{\includegraphics[width=.7\textwidth]{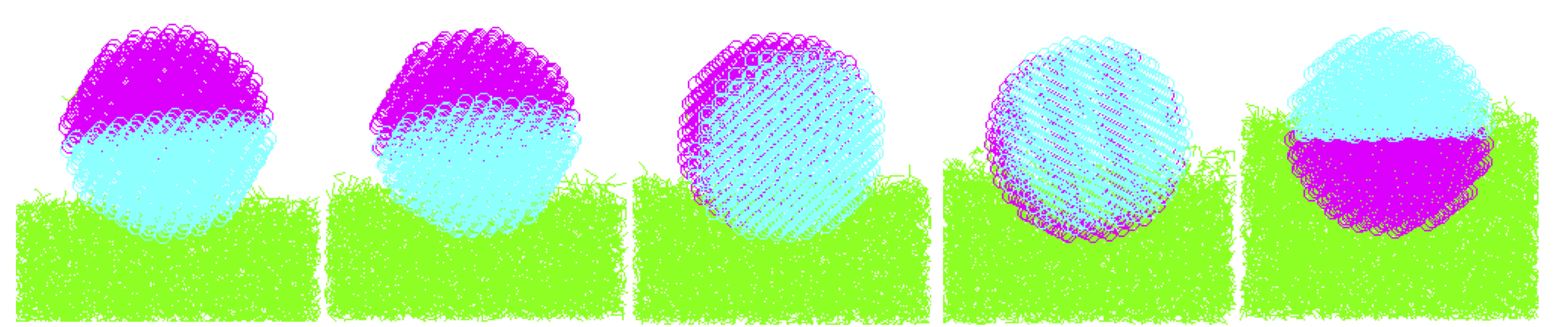}}
\subfigure[]{\includegraphics[width=.7\textwidth]{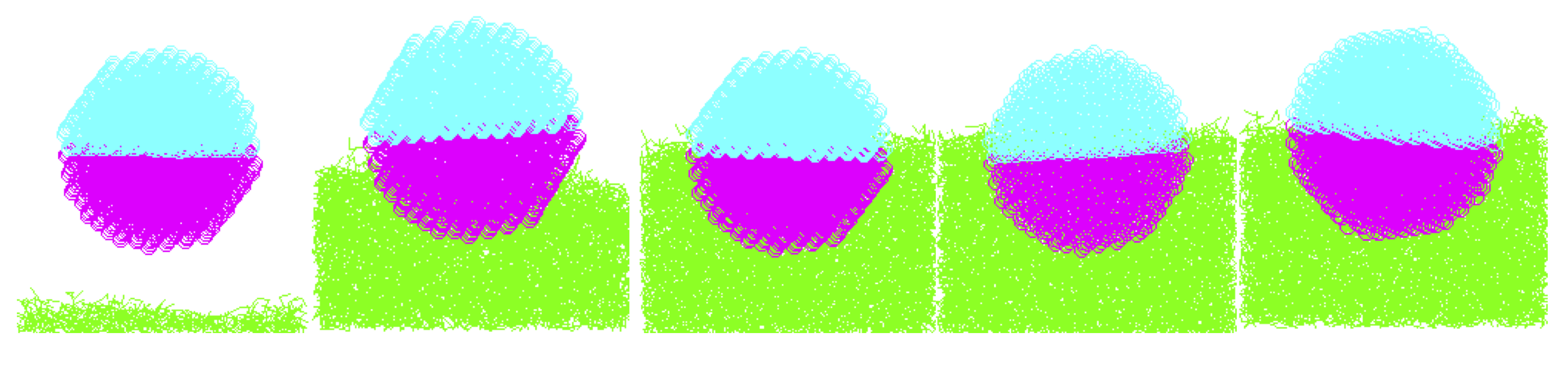}}
\caption{ Molecular free energy surface and numerical experiments for a 1.2/0.7 Janus particle. (a) Free energy surface, computed from thermodynamic integration.   (b) Particle released at 0$^\circ$ in vapor and not allowed to rotate, at times 0, 500, 1000 and 5000 $\tau$,  (c) Particle released at 0$^\circ$ in liquid and not allowed to rotate, at times 0, 500, 1000 and 5000 $\tau$. The last frame shows the configuration at the same final time when the particle is free to rotate. (d) Particle released at 0$\circ$ in vapor with rotation allowed, at times 500, 700, 800, 900 and 2500$\tau$ and  (e) Particle released at 180$^\circ$ in vapor and allowed to rotate, at times 0, 300, 600, 1000 and 5000$\tau$.}
\label{mdenergy7}
\end{figure*}

The continuum free energy landscape for the 1.2/0.8 Janus nanoparticle is
quite similar in shape, although the numerical values differ. The fact that
the interaction of the nonwetting face with the liquid is more favorable
($\gamma_{NW,L}-\gamma_{NW,V}$, is equal to 0.313 instead of 0.554 for the
1.2/0.7 particle)  allows the landscape to lie below that of the 1.2/0.7
particle for all orientations and immersions in which the nonwetting phase
is in contact with the liquid. In particular, the global maximum in the
energy for the configuration in which the nonwetting phase is in complete
contact with the liquid  ($\theta=0$ and $z/\sigma=38.3$) is now negative,
$2 \pi R^{2} (\gamma_{NW,L}-\gamma_{NW,V})-\pi R^{2} \gamma_{f}=-8.4
\epsilon$ and the interfacial energy of the completely immersed particle is
$2 \pi R^{2} \left
[(\gamma_{W,L}-\gamma_{W,V})+(\gamma_{NW,L}-\gamma_{NW,V}) \right ]$ =
$-119.0 \epsilon$. Note that the energy at the preferred configuration for
the 1.2/0.8 particle remains the same as the 1.2/0.7 colloid as only the
wetting face is in contact with the liquid ($-379.2 \epsilon$), and the two
energy minimum are retained for $\theta =0$ although the position of the
minimum in the vapor phase is now shifted to $z_p=38.3\sigma
+R(\gamma_{NW,L}-\gamma_{NW,V})/\gamma_f =42.0\sigma$ 

\subsection{Molecular landscape}

\begin{figure}
\centering
\includegraphics[width=0.3\textwidth]{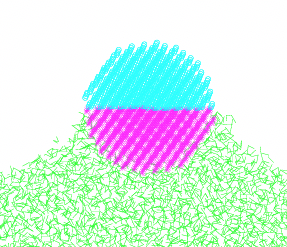}\hspace{0.1in}
\includegraphics[width=0.3\textwidth]{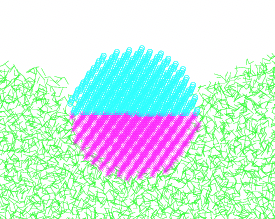}\hspace{0.1in}
\includegraphics[width=0.3\textwidth]{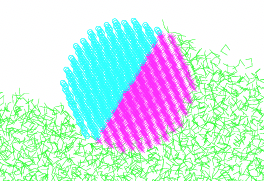}
\caption{Deformations of the liquid/vapor interface in the presence of a 1.2/0.7 Janus particle. Intermediate states during thermodynamic integration at (a) $z_p=x$ and $\theta=180^\circ$, (b) $z_p=x^\circ$ and $\theta=180^\circ$, (c) $z_p=38\sigma$ and $\theta=120^\circ$.}
\label{deformed}
\end{figure}

The interfacial energy landscape ($\Delta {\cal{F}}^{J,MD}$ of a Janus
nanoparticle, taking account of atomic scale roughness and variations in
meniscus shape, is shown in Fig. \ref{mdenergy7}a. The calculation uses the
molecular dynamics model (Fig. \ref{setupa}) and the thermodynamic
integration algorithm of eq.~\ref{fe2} for the 1.2/0.7 particle. Here, for each $\theta$ we integrate in $z$ from the vapor phase to deep into the liquid.  The landscape displays the same general features as the continuum version (Fig. \ref{continuumenergy}b) with the preferred configuration retaining a global minimum and the inverse configuration, with the nonwetting face completely wetted, having the largest energy.  These global minima and maxima are not as sharp as in the continuum case, and the relative smoothness of the MD topology relative to that of the continuum is a general feature of the MD result. This raggedness in the MD case results from force and interfacial shape fluctuations in the calculation, and is unavoidable in the absence of unlimited computer time. There is a {\em systematic} difference between the MD and continuum surfaces arising from deformations in the shape of the liquid/vapor interface due to the particle.
In Fig.~\ref{deformed} we show two snapshots of the particle and nearby
interface during the free energy calculation at 180$^\circ$ (only the atoms
in a slab-shaped region of width 10$\sigma$ about the midplane of the
particle areshown). When the particle center is above the original flat interface the liquid rises on the wetting surface up to the Janus boundary and when the center is below the original interface the interface falls to follow the boundary.  At 120$^\circ$, to give another example, the interface to the right coats the particle up to the Janus boundary but on the left has the appropriate contact angle ($132^\circ$) for the 0.7 side. This figure also shows clearly that the surface wetting along the two Janus faces drives the formation of a meniscus around the particle, and that the energy minimization at each value of $z_p$ and $\theta$ in the MD landscape is a balance between surface equilibration (which tends to reduce the interfacial energy) and meniscus formation (which tends to increase the energy due to an increase in liquid interface area). In the continuum landscape, which does not allow surface equilibration and the deformation of the liquid interface and computes the interfacial energy by summing the contact surface energies of the Janus faces (eq.~\ref{fe1}), the contact angles are determined by geometric intersection of the flat surface with the contact line. This lack of competition underlies the sharp features of the landscape as the intersection passes across the Janus boundary. 

The most obvious differences between the continuum and molecular free energy
surfaces are the corrugations parallel to the $z$-axis in the MD case, a
numerical artifact whose origin is as follows.  Each force measurement at a
given $(z_p,\theta)$ is a time average in a noisy system, and has some
statistical error associated with it. The free energy at that point is an
integral over all previous values of $z$ starting in the vapor, and includes
the sum of the previous point-wise errors, and since there is a time delay
in accumulating force values between different points the force errors are
statistically independent. Therefore, at each $\theta$ the error is the sum
of uncorrelated random variables and is analogous to a random walk.  The
error at each $\theta$ would then grow as the number of points, proportional
to $(z_p-z_0)^{1/2}$, but with a prefactor that would vary from one value of
$\theta$ to another, since the different values correspond to different
calculations. The corrugations thus represent different random walk
fluctuations superposed on the correct surface along each measurement line
in $z$. The strength of the fluctuations could be reduced by using a longer
averaging interval to lower the error associated with each force
measurement, but cannot be eliminated. The direction of the corrugations
could be altered by another choice of integration protocol, increasing with
decreasing $z_p$ if we had started in the fluid rather than the vapor, or
running parallel to  the $\theta$ axis if we had integrated in that
direction. Practically speaking, since the fluctuations increase with depth,
the molecular surfaces are most reliable at large $z$ but deteriorate moving
from right to left. In particular, the free energy results should be
independent of $\theta$ once the particle is well below the surface, as in
the continuum version, but the noisy molecular calculation misses this
feature. Specifically, if we average over the final free energy values in
$\theta$ at the end of the sequence (the lowest $z_p$) we obtain
-$(1.89\pm 10.4)\epsilon$ for the
1.2/0.7 case instead of -22.1 and -$(84.3\pm 10.2)\epsilon$ in the 1.2/0.8
case instead of -119,
showing some degree of convergence.  These discrepancies should be viewed in
the context of long-time excursions of simple random walks, where
convergence is often slow \cite{feller1968}.

Corrugations aside, the differences between the continuum and MD free energy surfaces have a clear physical origin.  At $\theta=0^\circ$, the MD free energy lies below the continuum energy but both of the continuum minima are present, although somewhat broadened.  Proceeding from the point at which the nonwetting face of the Janus particle breaches the interface from the vapor side to where the center of the particle is just above the interface ($46.3 \sigma <z_{p}< 40\sigma$), snapshots of the meniscus around the particle show that the interface is relatively flat as the intersection of the flat surface with the nonwetting face leads to obtuse angles. As in the continuum case, the interfacial energy lowers with $z_p$ due to th decrease in the area of the fluid interface which offsets the increase in energy due to the contact of the liquid with the nonwetting side of the Janus particle. The fact that the MD energies are lower can be attributed to the equilibration with the surface, which is more important than the surface deformation. As the center of the particle comes to the interface location ($z=38.3 \sigma$), the wetting face comes in contact with the liquid and snapshots show that the liquid tends to wet the entire hemisphere for $z_p < 38.3 \sigma$, creating a rising meniscus.  Although this wetting interaction with the surface lowers the energy (as it does for the continuum case) the cost of the deformation does not allow for the sharp reduction in energy as is evident in the continuum calculation. As the particle becomes completely immersed in the liquid the energy should return to the flat continuum value but falls below due to the numerical artifact described above, although the minimum reflecting the equilibrium of a homogeneous particle with the wetting interaction face (1.2) is recovered.

\begin{figure*}
\centering
\subfigure[]{\label{}\includegraphics[width=0.6\textwidth]{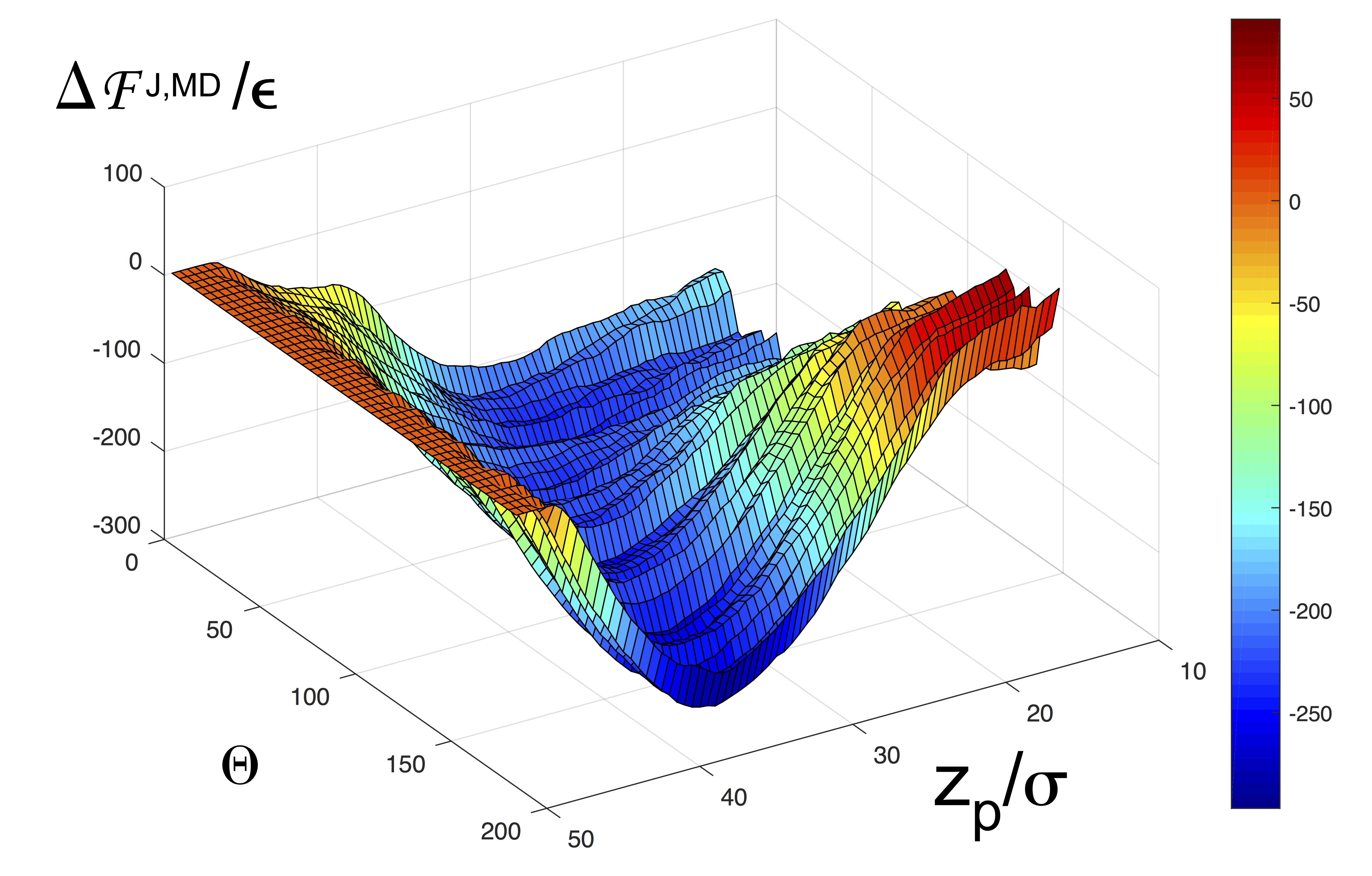}}
\subfigure[]{\label{}\includegraphics[width=0.7\textwidth]{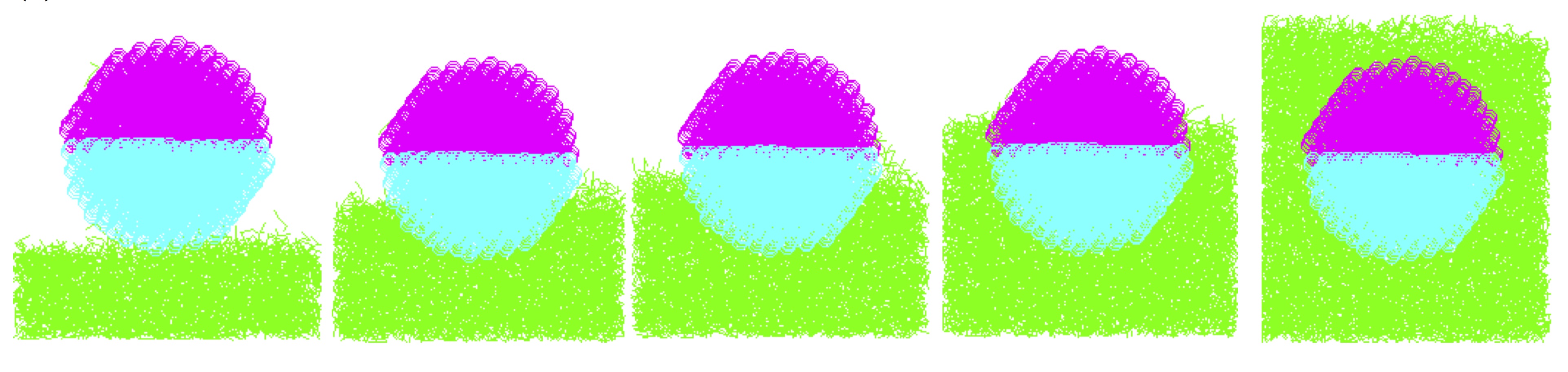}}
\caption{ Molecular free energy surface and numerical experiments for a 1.2/0.8 Janus particle. (a) Free energy surface computed from thermodynamic integration. (b) Particle released at 0$\circ$ in vapor and not allowed to rotate, at times 300, 500, 800, 1000 and 5000 $\tau$}.
\label{mdenergy8}
\end{figure*}

For $\theta=180^{\circ}$ snapshots of the meniscus formation around the
particle as the particle becomes immersed in the liquid wetting side down
show that as even before the particle touches the nominal interface ($z_p
\approx 46.3 \sigma$), fluctuations in the liquid begin to form a wetting
cap at the bottom of the particle (see Fig.~\ref{deformed}), creating a
rising meniscus which wets the particle up to the Janus boundary long before
the boundary drops to the original meniscus height. The wetting leads to a
reduction in the interfacial energy as in the MD case.  Similar interfacial
deformations for a partially immersed homogeneous particle have been
observed, presumably due to contact-line pinning at heterogeneities
\cite{butt2018}. Recall that in the continuum case, the energy is
also reduced because of the reduction in the area of the liquid interface.
In the MD case, while the liquid wets the wetting hemisphere, an energy
penalty arises because the rising meniscus creates a larger interfacial
area. As a result, the MD curve lies above the continuum curve. As the
particle descends further into the liquid and its center approaches the interface location ($46.3 \sigma <z_{p}< 38.3\sigma$), the meniscus becomes flatter and the interfacial energy decreases until the global energy minimum is reached at $-232 \epsilon$, which is much larger than the continuum value $379.3 \epsilon$. Although the interface in the MD case is relatively flat when the Janus boundary sits on the interface, the integrated effect of the meniscus deformation on the path from the vapor phase to the position of half immersion causes the free energy reduction to be less than in the continuum case.  With further immersion into the liquid $z_p < 38.3 \sigma $, the interface become pinned at the contact angle causing an increase in interfacial area (and an increase in interfacial energy). Eventually, the meniscus detaches form the Janus boundary, and the liquid wets the nonwetting face with a descending meniscus intersecting at an obtuse angle with the surface and a less deformed meniscus.  As in the continuum case, the interaction of the liquid with the solid raises the interfacial energy, but it does not plateau to the continuum value, as discussed above.

The interfacial energy landscape for the 1.2/0.8 Janus particle in 
Fig.~\ref{mdenergy8}a shows the same general features as the continuum
version in Fig. \ref{continuumenergy}c, as was the case previously. The
1.2/0.8 MD landscape lies below the 1.2/0.7 landscape for all situations in
which the liquid is wetting the nonwetting face, due to the stronger
interaction of the liquid with the 0.8 face. The MD minimum energy, achieved
when the wetting face is completely immersed, is approximately the same for
the 1.2/0.8 and 1.2/0.7 Janus particles (Fig. \ref{onedlandscape}), as was
the case with the continuum profiles, since the particle interaction with
the liquid is only through the wetting face with a common interaction
parameter (1.2). The most significant difference between the continuum and
the MD landscapes occurs along the $\theta = 0^\circ$ curve (Fig.
\ref{onedlandscape}), where the minimum characterizing a homogeneous
particle with a 0.8 face is not realized, as it is for the continuum case at
$z_{p}/\sigma=42$ (and for the 1.2/0.7 case).  Snapshots show that in this range of $z_{p}$ the stronger interaction of the nonwetting face with the liquid allows the liquid to climb over the Janus boundary and reach the wetting hemisphere, giving a reduction in the energy (although not asymptoting to the continuum value for complete immersion).

\section{Hysteretic Phenomena in the Attachment of Janus Particles to the Interface}

The MD interfacial energy landscape constructed by thermodynamic integration provides a basis for understanding how Janus colloids will situate and orient themselves when attaching to the vapor/liquid interface from either the vapor or liquid phases bounding the interface. In particular, the strong heterogeneity of the Janus particle which generates for the local minima in the landscapes may give rise to multiple metastable equilibrium orientational states. To explore this possibility in a context related to experimental practice, we undertake MD simulations in which the colloid is positioned either above the interface in the vapor phase, or below the interface in the liquid phase and allowed (due to a small downward or upward velocity) to come to the proximity of the surface. The particle is then allowed to diffuse to the surface, attach, orient, and migrate along the surface due to diffusion. We aim to understand the motion in terms of trajectories on the energy surface $\Delta {\cal{F}}^{J,MD}(z_p,\theta)$

Consider first a 1.2/0.7 Janus particle released from the vapor phase at $t=0$, at an orientation $\theta = 0^{\circ}$ (nonwetting side down) and allowed to migrate to the surface by the intermolecular interaction of the atoms of the particle with the liquid phase.  To isolate the effect of rotation in finding an equilibrium in the energy landscape, the orientation is at first fixed (no rotation or NR) and the resulting motion is shown in snapshots at $t$=0, 500, 1000 and 5000$\tau$ in Fig. \ref{mdenergy7}b. in this case, the particle follows a trajectory on the one dimensional energy curve $\Delta {\cal{F}}^{J,MD}(z_p,0^\circ)$ plotted in Fig.~\ref{onedlandscape}a, and as it is not allowed to rotate it does not migrate to the global minimum at $\theta=180^{\circ}$, but instead diffuses to the first minimum at $z_p/sigma=40$.
As we noted, this minimum characterizes the minimum energy of a homogeneous particle with a $0.7$ interaction attached to the surface, and this is exactly what the MD simulation shows as the particle sits at this minimum with only its nonwetting face immersed in the liquid.  The fluid near the interface fluctuates and a molecule or two even climbs up the strongly wetting side, but this configuration is stable over the duration of the simulation. Thus the dipolar nature of the Janus particle does not effect the equilibrium attachment orientation when rotation is not permitted and the particle is inserted from the vapor phase non-wetting side down. 

Compare this behavior with the case in which the particle is released from the \textit{liquid} side ($z_p/\sigma = 30$) with the same fixed orientation: see Fig. \ref{mdenergy7}c. This particle is released within the broad minimum of the one dimensional energy curve of Fig. \ref{onedlandscape}a and remains in this minimum. The strong interaction of the liquid with the wetting phase allows the particle to remain completely immersed. Comparison of the trajectories and simulations for these later two cases demonstrates that when the Janus particle is unable to rotate, the particle experiences a significant hysteresis: When inserted (wetting side down) from the vapor it comes to an equilibrium ``high'' on the interface with its nonwetting side in contact with the liquid. When inserted (wetting side down) from the liquid, it remains in the liquid phase, with the strong interaction of the wetting face with the liquid keeping it immersed. Once immersed in the liquid phase, if allowed to rotate it remains there: the final snapshot ($t=5000\tau$) in Fig. \ref{mdenergy7}c is taken if rotation is permitted from the same starting point in the liquid. With rotation, the particle should migrate to the global minimum by flipping over, but for this completely immersed particle, although free to rotate the relevant rotational diffusion time scale is long, roughly 10$^5\tau$ for a homogeneous sphere of this size \cite{koplik2017}, and well beyond the duration of these simulations.

\begin{figure}
\centering
\includegraphics[width=0.3\textwidth]{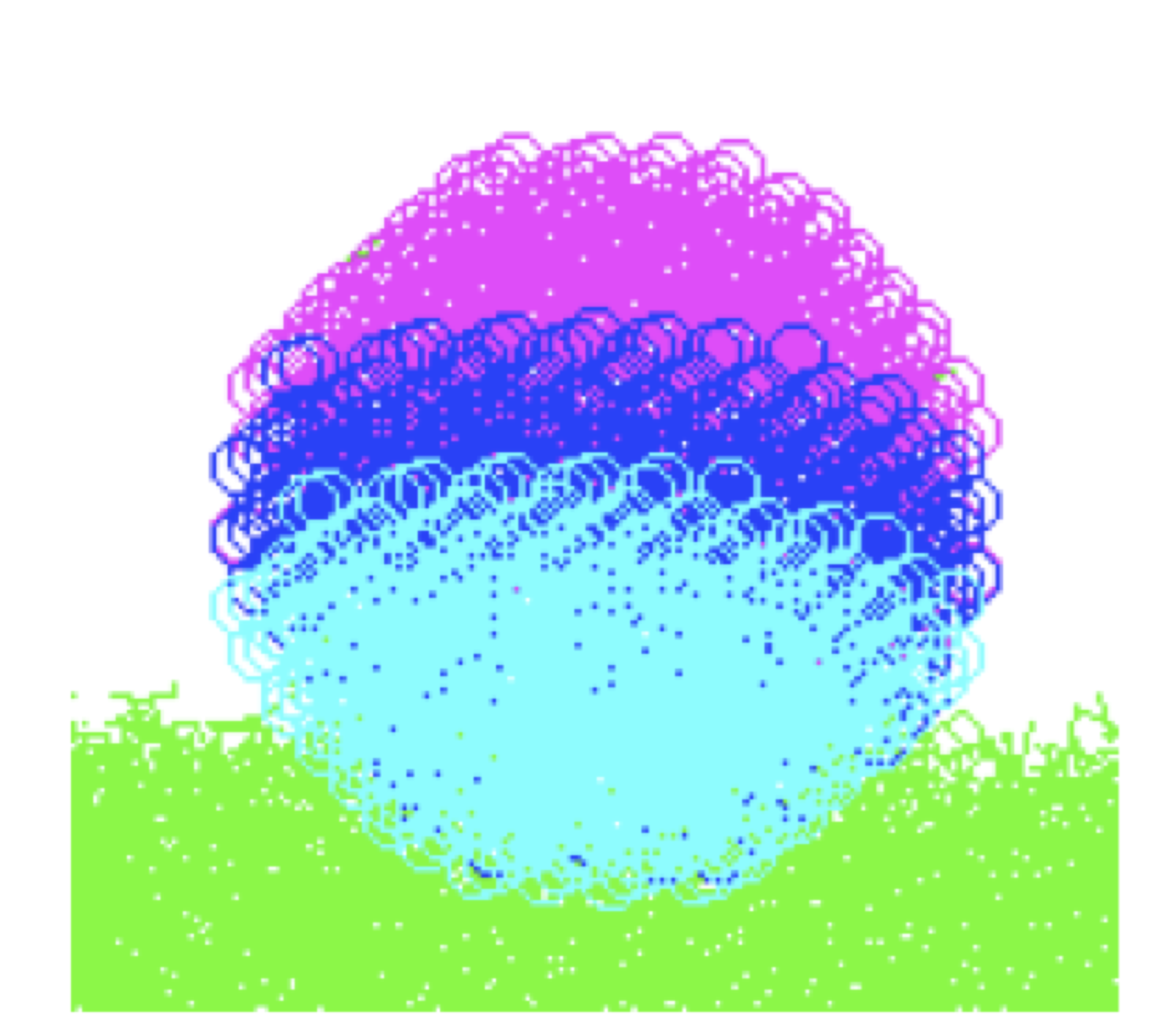}
\caption{A three banded particle with a small central wettability.}
\label{threeheads}
\end{figure}

Contrast the above behavior without rotation to the case when the particle is allowed to rotate freely and is released from the vapor side with the nonwetting side down, shown in Fig.~\ref{mdenergy7}d.  It initially halts due to the weak wettability of the part of the particle it contacts and the relative flatness of the landscape near the local (MD) minimum at $\theta=0^{\circ}$ and $z_{p}/\sigma \approx 42$, but orientational fluctuations allow fluid molecules to reach the nearby part of the strong side, resulting in a torque which flips the particle over. Once the strong side is fully immersed the particle sinks to the equilibrium depth of the global minimum with the Janus boundary at the interface (Fig. \ref{mdenergy7}d.  One might imagine that if the wettability of the weak side is reduced further the particle's downward motion would arrest at a higher point and prevent fluctuating fluid molecules from reaching the strong side. Unfortunately, in this case (1.2/0.6) the particle is in poor contact with the liquid and has more freedom to rotate, and it still tips over. At still lower weak wettability values the particle simply bounces between the interface and the top of the simulation box.  However, we observe that a three-banded Janus particle (Fig. \ref{threeheads}) with a small central wettability ($c_{12}=1.2$ for  polar angle $0\le \pi/3$, $c_{12}=0.5$ for $\pi/3\le 2\pi/3$ and $c_{12}=0.7$ for $2\pi/3\le \pi$) will arrest at the interface and  maintain a orientation near 0$\circ$ for 10,000$\tau$. 

When the initial orientation of the particle is wetting side down ($\theta^{\circ}=180$) the migration with and without rotation is straightforward. Fig.~\ref{mdenergy7}e shows the migration of a particle released in the vapor phase wetting side down with rotation allowed. Aside from fluctuations it settles directly into the minimum free energy state with the strong side immersed within $2500\tau$.  If instead the particle is released with the same orientation below the interface with or without rotation allowed (not shown), it immediately rises to the global minimum.  The distinction between this case and Fig.~\ref{mdenergy7}c is that there the free energy surface is rather flat, corresponding to the long rotational diffusion time, but here the free energy surface slopes sharply towards the global minimum (see Fig.~\ref{onedlandscape}b).   

For the 1.2/0.8 Janus particle, we noted above that the one major difference in the energy landscape was the loss of the minimum for the one dimensional landscape at $\theta =0^{\circ}$, due to the stronger interaction of the nonwetting side with the liquid for the 1.2/0.8 particle relative to the 1.2/0.7 particle. For the 1.2/0.7 Janus particle, the presence of this minimum allowed the particle, without rotation, when inserted from the vapor side (nonwetting side down) to come to a metastable equilibrium in which the nonwetting side is partially immersed in the liquid. In the absence of this minimum, the 1.2/0.8 Janus particle cannot obtain such a metastable equilibrium and as shown in the snapshots of Fig.~ref{mdenergy8}b for $t$=300, 500, 800, 1000 and 5000$\tau$, the particle becomes fully immersed in the liquid. The weak side of the particle has a slightly stronger attraction to the liquid so the particle initially arrests at a slightly deeper depth than in the 1.2/0.7 particle, cf. Fig. \ref{mdenergy7}b . The strong side of the particle is closer to the liquid, so it is more easily reached by fluid molecules fluctuating above the average interfacial position, which are then pinned there. Each such pinned molecule attracts others, forming a thin film on the particle surface and pulling it down into the liquid. Once the upper side of the particle is in contact with denser liquid the strong attraction pulls it in. The final depth is the same (up to fluctuations) as when the particle is released below the interface in (b), and there is no hysteresis.  (Strictly speaking we should compare this case to a 1.2/0.8 particle released below the interface, but there is little difference in the final depth.)  

To summarize the results of these migration studies, we observe three possible final states rather than a single minimum free energy state, and these arise from a mixture of imposed constraints and time limitations. The behavior in case Fig.~\ref{mdenergy7}b arises because an explicit constraint prevented rotation and a simulation time limitation prevented the occurrence of a large vertical fluctuation. Likewise, the final states in Figs.~\ref{mdenergy7}c and \ref{mdenergy8}b do not reach the global minimum due to the long time needed for a fully immersed particle to rotationally diffuse through a large angle so that the weak side senses the interface and is expelled.

\begin{figure*}
\centering
\includegraphics[width=.17\textwidth]{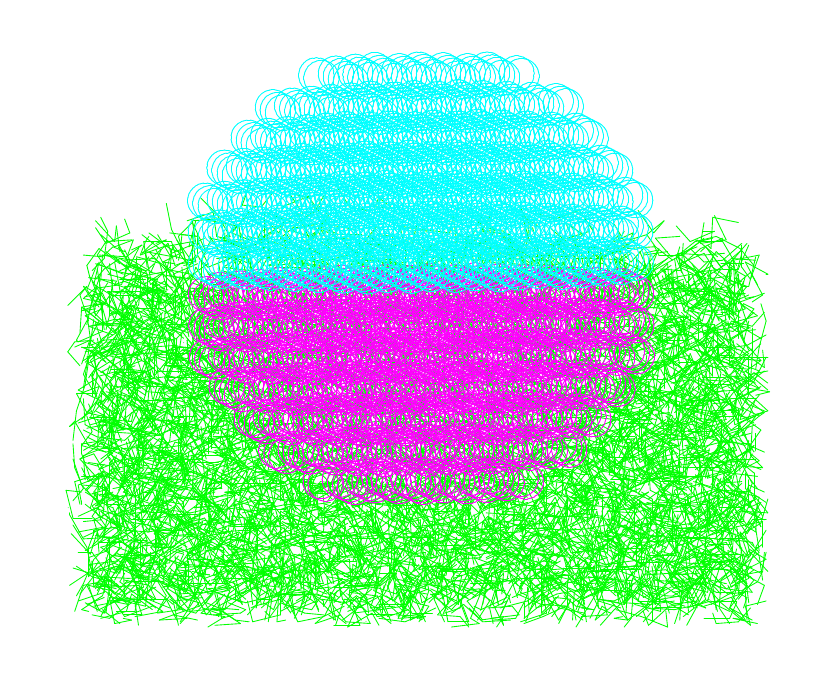}
\includegraphics[width=.17\textwidth]{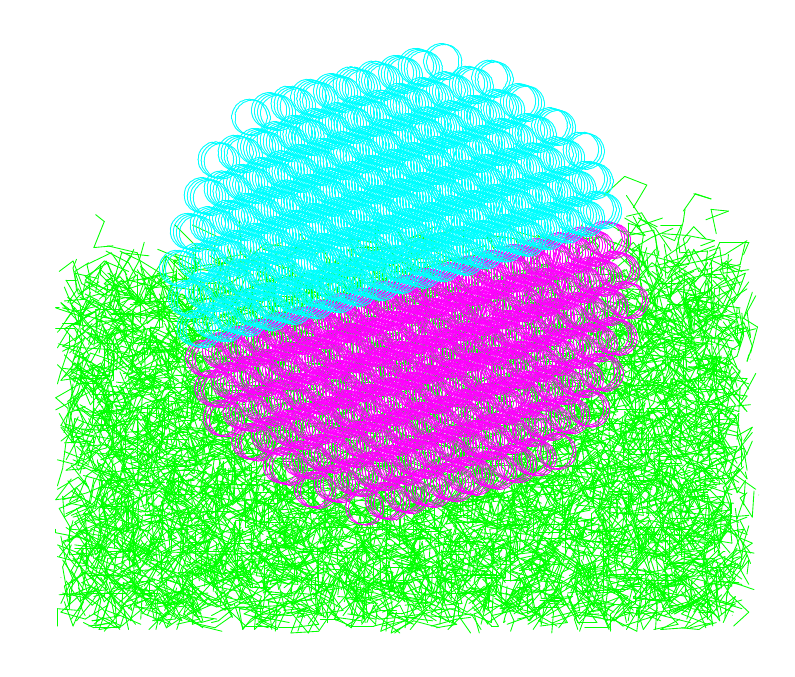}
\includegraphics[width=.17\textwidth]{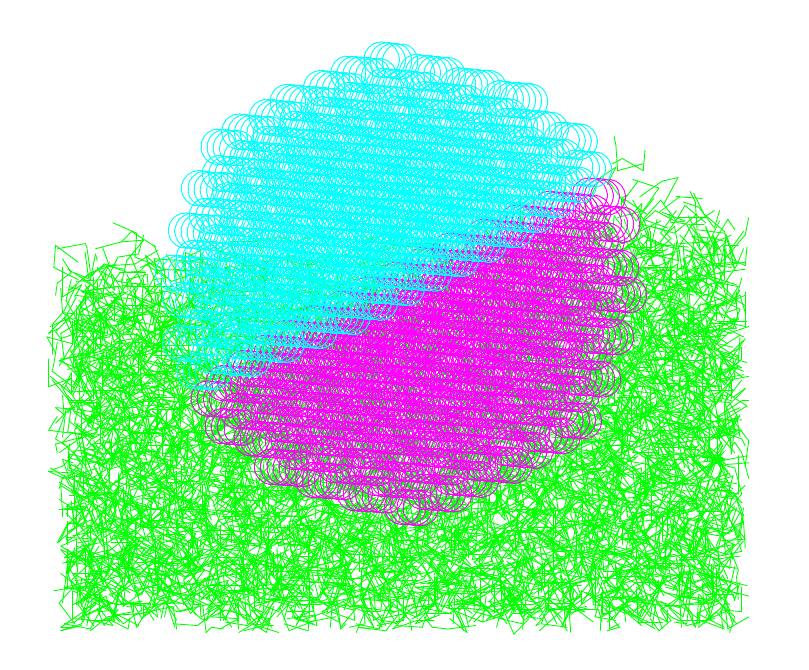}
\includegraphics[width=.17\textwidth]{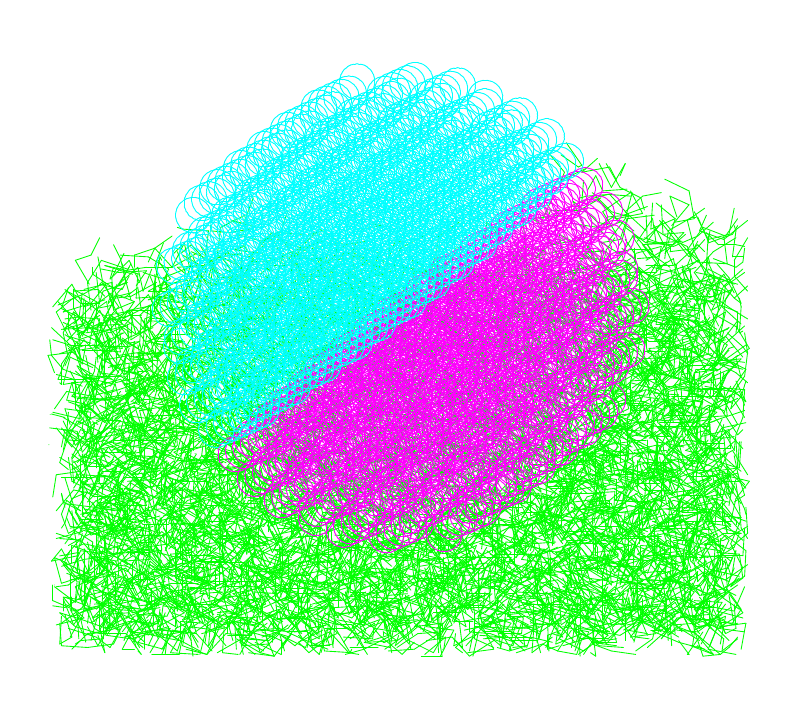}
\includegraphics[width=.17\textwidth]{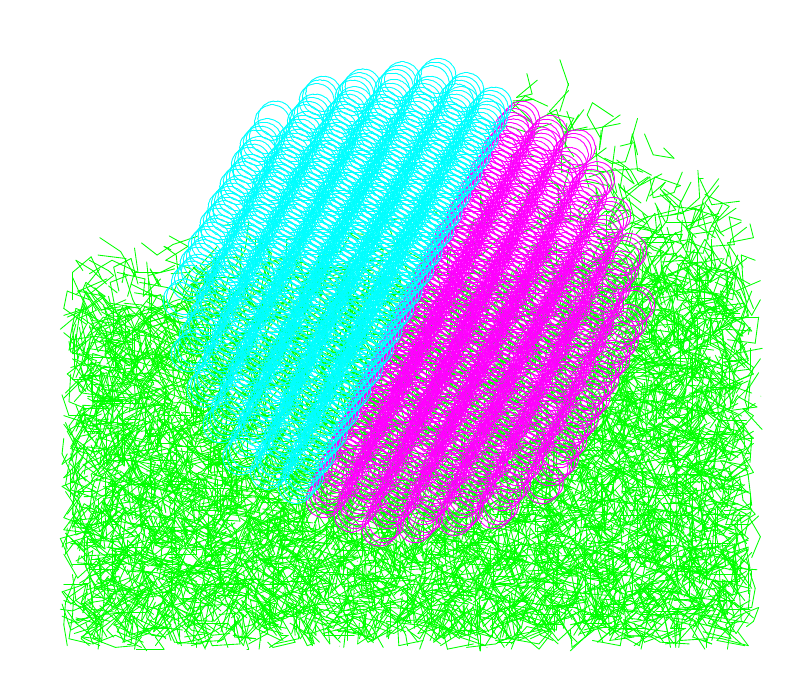}
\caption{Top: Snapshots of Janus spheres pulled from right to left along 
an interface at speeds (left to right) 0.01, 0.025, 0.5, 0.075 and
0.1$\sigma/\tau$.}
\label{pull} 
\end{figure*}

\section{Drag and Pinning of Janus Particles During Motion Along The Surface}
\label{drag}

\begin{figure}
\centering
\includegraphics[width=.3\textwidth]{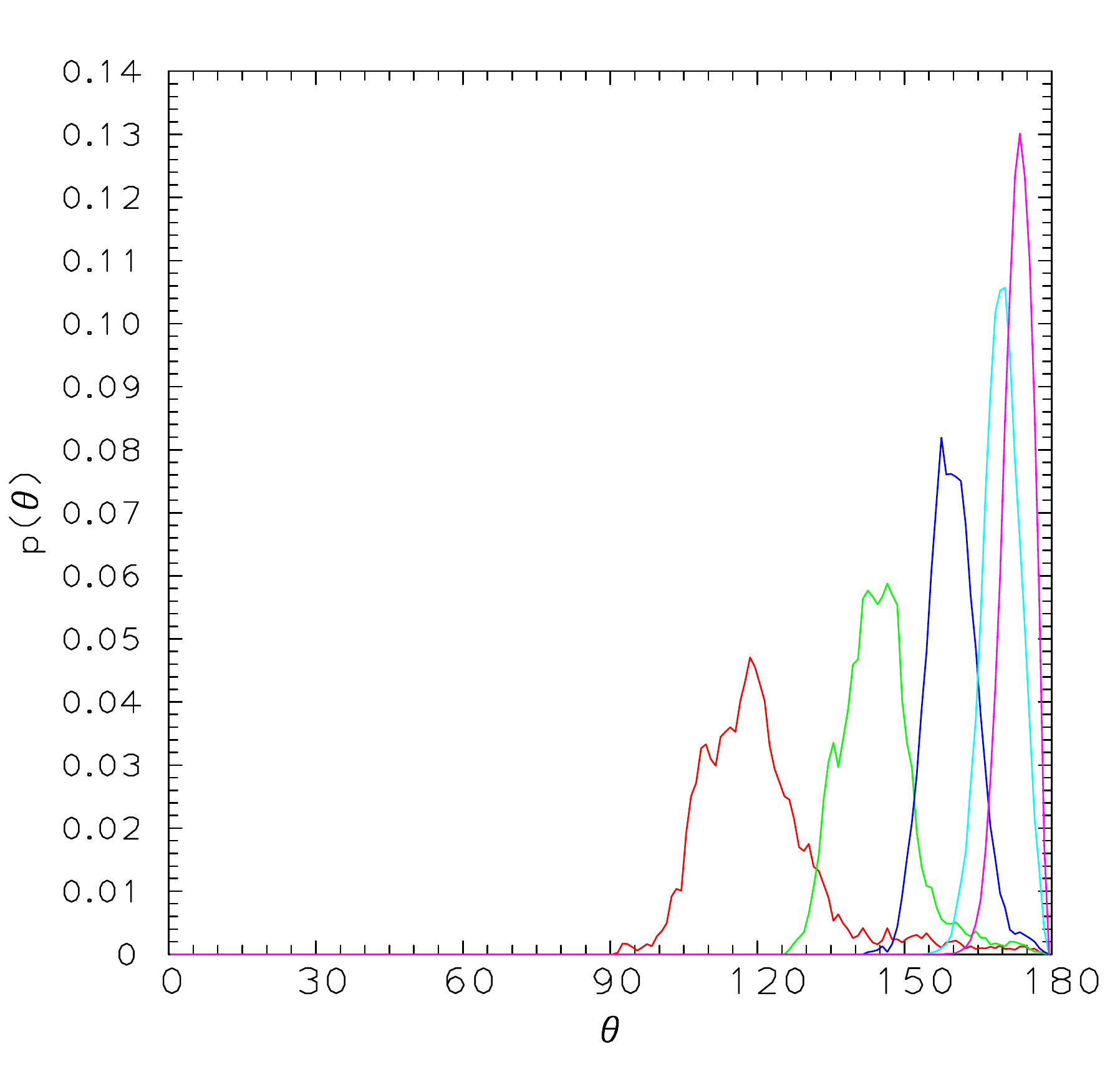}
\caption{Top: Probability distribution of particle orientation when pulled
right to left along an interface at speeds 0.01 (magenta), 0.025 (cyan), 0.5
 (blue), 0.075 (green) and 0.1$\sigma/\tau$ (red).}
\label{orient}
\end{figure}

In this section, the translation and rotation of Janus particles attached to an interface is studied.  As in KM where MD simulations of homogeneous particles with nanoscale roughness at an interface were studied, the center of the particle is fixed at the average equilibrium depth and translated at constant velocity parallel to the 
interface, and the force $F$ exerted on the particle by the surrounding fluid
atoms is computed.  For homogeneous spheres results were reported at a single
pulling velocity, $U=0.1$, but in fact the drag coefficient $\xi=F/U$ was
independent of pulling speed over a wide range (but see below).  For Janus 
spheres, however,  we see a weak but
systematic variation:  the drag coefficients for each velocity is recorded
in Table I.  The simulation results are an average over a statistical
ensemble, where the number of independent realizations (different random seed 
used to generate the initial velocities) is larger at lower velocities to
reduce the statistical error,  At still lower speeds the fluctuations 
are too strong to reliably extract the force, while at higher values the
interface becomes distinctly non-planar away from the particle and cannot be
sensibly compared to planar cases. 
Along with the drag variation there is a systematic change in
the particle's orientation, defined here by a director vector
fixed in the particle running from the center of the high-wettability side
to the opposite pole.  In Fig.~\ref{pull} we show typical snapshots of the
system in the steady state at various values of $U$ (the particle moves
to the left). Of course the particle orientation fluctuates in this
situation, and in Fig.~\ref{orient} we show the corresponding 
pdf of the angle between the director and the vertical.

\begin{table}[h]
\begin{tabular}{|l||l|c|c|} \hline\hline
$U$   & $\xi$      & $ <\theta>$ & $N$ \\ \hline
0.1   & $386\pm 5$ & $122\pm13$  & 10  \\ 
0.075 & $384\pm5$  & $145\pm8$   & 15 \\ 
0.05  & $386\pm7$  & $161\pm6$	 & 20 \\ 
0.025 & $386\pm10$ & $171\pm4$   & 40 \\ 
0.01  & $392\pm14$ & $174\pm3$   & 99 \\ 
0.0   & $397\pm5$  & $173\pm4$   & 90 \\ \hline 
\end{tabular}
\label{table1}
\caption{Drag coefficient (units: $m/\tau$) for Janus spheres pulled along
the liquid/vapor interface at various velocities, along with the average 
director orientation and the number of realizations $N$ used in the 
calculation. $U=0$ means pure diffusion.}
\end{table}

We also compute the Janus particle diffusivity, by placing it at the 
interface and allowing it to diffuse freely for 10,000$\tau$. A statistical
ensemble of 90 realization is used, obtained by choosing different random 
number seeds to generate the initial velocity distribution.  During the
simulation the particles are allowed to translate and rotate freely, but in 
fact there is little motion normal to the interface and little rotation. 
If we denote the
particle's position as $({\bf r}(t),z(t))$, then the mean-square fluctuation
in vertical position is $\langle\Delta z^2\rangle=0.548\sigma^2$. Likewise, 
the mean square fluctuation in orientation angle relative to the
vertical is 4.1$^\circ$. We obtain the diffusivity using 
$\langle{\bf r}^2(t)\rangle)\sim 4Dt$, and use the Stokes-Einstein relation to find the
drag coefficient $\xi=k_BT/D=(397\pm 5)m/\tau$.  The results is consistent
with the pulled drag at the lowest velocity but deviates as the velocity
increases. 

To place the numerical value of the drag in perspective, we note that 
the Stokes drag on a fully immersed sphere in the simulated 
fluid used here is $6\pi\eta R = 781 m/\tau$ theoretically, and 698, 756
and 853 $m/\tau$ for $c_{12}=$0.8, 1.0 and 1.2, respectively,
in MD simulations in this system using a fully immersed homogeneous sphere. 
The drag increases with $c_{12}$ because increasing the wettability draws
fluid molecules closer to the sphere and effectively increases its radius.
Alternatively, one may say that increasing wettability decreases slip.

The physical origin of the drag variation is the presence of two
competing effects. When a particle is pulled along an interface there is more
fluid resistance on the part of surface that is immersed than on the part in
vapor, producing a torque which tends to rotate the particle. (This effect
is incorporated in the resistance tensor formalism\cite{kim1991microhydrodynamics} ,
in which force and torque are coupled to linear and angular velocity.) 
If, as in the figure, the particle translates from right to left, the torque
is counterclockwise.  However, this rotation tends to withdraw the strongly
wetting part of the particle surface from the liquid, which then forms an
irregular coating film on the top of the particle. There is then an
unbalanced force at the contact line, or equivalently an increase
in the area of the liquid/vapor interface, which exerts a clockwise torque.
In addition, the same rotation places less wettable regions of the 
particle surface in contact with dense liquid, which reduces the drag force and the drag-induced 
torque.  In the simulations, the interface is flat in equilibrium and when the
particle is translated it initially experiences a counterclockwise torque
in response to the drag imbalance. The rotation progressively entrains a 
surface film with a clockwise torque, and at the same time reduces the   
other torque, and the particle reaches a steady state when the torques
balance. The magnitude of the effect is set by the pulling velocity, so the
deviation of the orientation from the vertical increases with speed.
Likewise, the change in drag coefficients is a competition between the
increase due to arresting the rotation and the decrease due to exposing 
less wettable surface to the liquid, and it happens that the latter effect
dominates.

The conclusion to be drawn from these calculations is that a heterogeneous
Janus sphere changes its orientation when it is pulled along an interface
and as a result the drag coefficient will vary with pulling speed.
A particle pulled sufficiently rapidly has a strong tilt and a reduced drag
in comparison with a diffusing particle, and an apparent deviation from the
Stokes-Einstein relation will result.  In these simulations, the effect is
a rather weak drag reduction, and in particular there is no evidence for a
enhanced drag above the fully immersed value.  

\begin{figure}
\centering
\includegraphics[width=0.3\textwidth]{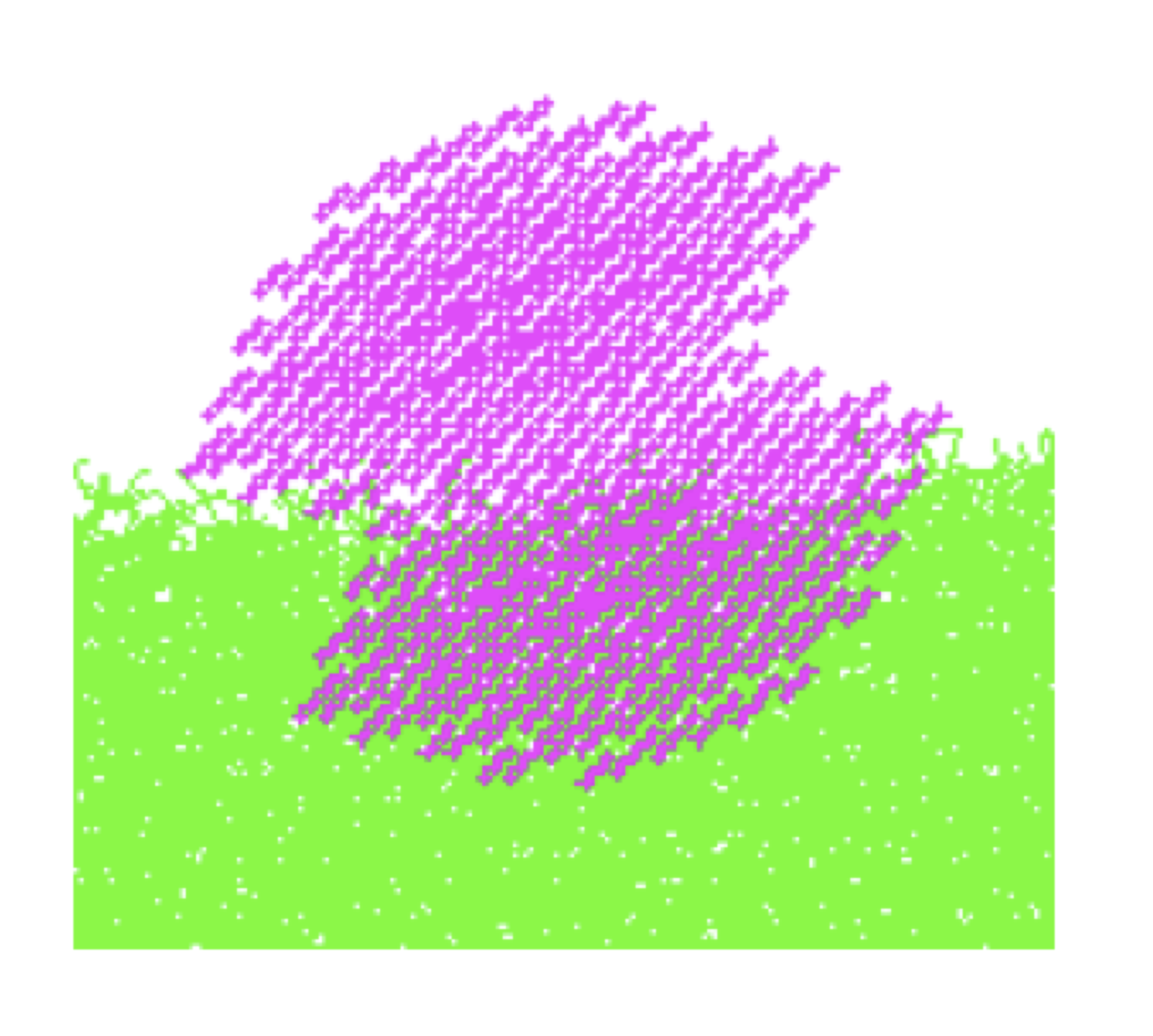}
\caption{Hourglass shaped homogeneous particle with atoms interacting with liquid with a single interaction $c_{12}$.}
\label{hourglass}
\end{figure}

The fact that the drag coefficient increases in the absence of rotation is 
quite general, and we illustrate this point with two examples. First, 	
an analog of the Janus sphere
having structural rather than chemical disorder is an hourglass-shaped
particle, shown in Fig.~\ref{hourglass}.  The particle is obtained by
selecting the atoms in a section of a cubic lattice: atomic positions satisfying
$(x^2+y^2)^{1/2}\le 6+2\cos{2\theta}$ in spherical coordinates.
In this case the particle has
uniform wettability, and as in KM the value of $c_{12}$ controls the
immersion depth.  We repeat the pulling calculations for different
wettability values, at two different velocities, with the results given in
Table II.  In this case the heterogeneity is not strong enough to pin the
particle at a fixed orientation in general, but the rotation is arrested 
for the lower velocity for particles with $c_{12}\le0.8$
that are less than half immersed in the liquid.
As seen in Table II, the drag coefficient is approximately speed-independent 
when free rotation occurs, but when the heterogeneity is strong enough to
arrest the rotation the drag coefficient is higher.  In this case the
variation in drag is considerably stronger, because there are no competing 
surface tension effects. Note again, that the
enhanced values of the drag coefficient are well below the fully immersed
value, since the latter should be even higher that those for
$c_{12}=1$.  The second example is a homogeneous sphere with uniform
interactions and only atomic-scale roughness, considered in KM.  In this case
the sphere rotates when pulled at all but the lowest velocity and
wettability, and again in that situation there is a distinct drag increase. 

\begin{table}[h]
\begin{tabular}{|l||c|c|c|c|c|}\hline
$U$\verb+\+$c_{12}$ & 0.6 & 0.7 & 0.8 & 0.9  & 1.0 \\ \hline\hline
\multicolumn{6}{|c|}{Hourglass}\\ \hline
0.1                 & 46.1 & 124 & 174 & 409 & 494 \\ \hline
0.03                & \textcolor{red}{74.0}  & \textcolor{red}{188} &
\textcolor{red}{200} & 413  & 487 \\ \hline\hline
\multicolumn{6}{|c|}{Sphere} \\ \hline
0.1                 & 33.6 & 88.9 & 187 & 320 & 428 \\ \hline
0.03                & \textcolor{red}{60.7}  & 90.6 & 197 & 326 & 419  \\
\hline
\end{tabular}
\label{table2}
\caption{Drag coefficient (units $m/\tau$) for hourglass and uniform
spherical particles
at two velocities (units $\sigma/\tau$) for various wettabilities. The
statistical errors are similar to those in Table I.  Red entries
indicate little or no rotation during the motion.}
\end{table}

\section{Conclusions}
We have used MD simulations to examine the role of heterogeneities in the
behavior of nanoparticles on a liquid/vapor interface.  
For the most part we focused on Janus spheres, having different wettabilities 
on each hemisphere, as a simple and tractable model.

One set of simulations explored hysteretic effects arising when a Janus 
sphere migrates toward an interface from bulk liquid or vapor phases. 
Simulations indicated that the particle could evolve into either of three 
possible equilibrium states, depending its initial position and orientation, 
with varying degrees of stability that could be explained in terms of the local
molecular interactions.  At the same time, we reinterpreted the observed behavior in a
more general manner, based on computations of the free energy surface - 
the Helmholtz free energy of the particle and fluid system as a function 
of particle position and orientation. As a byproduct of this analysis we 
noted that the standard  continuum calculation of the free energy of an
interfacial particle which does not take account of interfacial fluctuations
could be misleading when applied to nanoparticles, and that a molecular  
level calculation was required in that case.

A second set of simulations studied the drag and diffusion of Janus particles. When diffusing along
the interface these particles fluctuate weakly about a configuration with
the more wetting hemisphere in the liquid and the less wetting side exposed
to vapor. When pulled along the interface by an external force 
they rotate into a steady-state tilted configuration with 
a tilt angle increasing with velocity, placing more and more  
of the less-wetting hemisphere into contact with the liquid. 
The result was a weak but significant variation of drag coefficient 
with pulling speed.  Analogous simulations of a structurally 
heterogeneous hourglass-shaped particle indicated that the particle would rotate
continuously at high pulling speeds but the rotation would arrest at lower
velocities, thereby increasing the drag coefficient. Even a very weakly
heterogeneous particle - a homogeneous nanosphere with only atomic-scale
surface roughness - showed the same transition at sufficiently low
pulling speeds. The conclusion is a heterogeneous particle may adopt different 
configurations when diffusing than when pulled, and since the drag
coefficient depends on the configuration, an apparent violation of the 
Stokes-Einstein relation may occur.  As a byproduct we were able to test the
suggestion that heterogeneous particles at an interface may have a drag
coefficient greater than the bulk value when fully immersed in a liquid,
but found only drags smaller that the bulk value.

A key question is whether results for nanoparticles are relevant to larger,
micron-sized colloidal particles. Some specific features of
nanoparticle systems, such as the effects of the finite width of a
liquid/vapor interface, would likely play a minor role, while other 
phenomena such as the importance of hydrodynamic fluctuations and 
the appearance of metastable configurations with long relaxation times are 
already well known to be significant at micron scales.  One result we
have emphasized, which is obvious in hindsight but not widely appreciated,
is that a heterogeneous particle may adopt different spatial and
orientational configurations under the action of different external stimuli
and accurate modeling is needed to reflect this variation. The apparent
Stokes-Einstein violations discussed in Section \ref{drag} illustrate this
point but the same caution applies to the modeling of all hydrodynamic and 
surface interactions.  In analyzing the simulation results we have used both 
intuitive arguments based on molecular configurations and more quantitative
ones based on the free energy surface.  The systems considered here were
sufficiently simple for qualitative reasoning to succeed but for more
general and varied patterns of competing heterogeneities the quantitative
method employing free energy surfaces is more promising. 

%

\end{document}